\documentclass[twocolumn,epjc3]{svjour3}
\usepackage[dvipsnames]{xcolor}
\usepackage{arydshln, booktabs, cite, feynmp-auto, graphicx, hyperref, listings, multirow, newtxtext, newtxmath, nicefrac, relsize, slashed, xspace}
\hypersetup{colorlinks=true, linkcolor=ForestGreen, citecolor=ForestGreen, filecolor=ForestGreen, urlcolor=ForestGreen}
\journalname{Eur. Phys. J. C}
\newcommand{\orcid}[1]{\begingroup\href{https://orcid.org/#1}{\includegraphics[width=9pt]{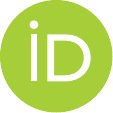}} \endgroup}
\DeclareUnicodeCharacter{2212}{-}

\def\madanalysis{{\lstinline{MadAnalysis}\,\lstinline{5}}\xspace}
\def\spey{{\lstinline{spey}}\xspace}

\newcommand{\be}{\begin{equation}}
\newcommand{\ee}{\end{equation}}
\newcommand{\bpm}{\begin{pmatrix}}
\newcommand{\epm}{\end{pmatrix}}

\def\met{E_T^\mathrm{miss}}
\def\ptmiss{\vec{p}_T^{\text{miss}}}

\newcommand{\ie}{\textit{i.e.}}
\newcommand{\etc}{\textit{etc.}}

\begin{document}
\title{Deciphering compressed electroweakino excesses with \madanalysis}

\author{
  Jack Y. Araz\,\orcid{0000-0001-8721-8042}\thanksref{e2,add5, add4,add2} \and
  Benjamin Fuks\,\orcid{0000-0002-0041-0566}\thanksref{e1,add1} \and
  Mark D. Goodsell\,\orcid{0000-0002-6000-9467}\thanksref{e3,add1} \and
  Taylor Murphy\,\orcid{0000-0002-3215-9652}\thanksref{e4,add3}
}

\thankstext{e2}{\href{mailto:j.araz@ucl.ac.uk}{j.araz@ucl.ac.uk}}
\thankstext{e1}{\href{mailto:fuks@lpthe.jussieu.fr}{fuks@lpthe.jussieu.fr}}
\thankstext{e3}{\href{mailto:goodsell@lpthe.jussieu.fr}{goodsell@lpthe.jussieu.fr}}
\thankstext{e4}{\href{mailto:murphyt6@miamio.edu}{murphyt6@miamioh.edu}}

\institute{Department of Physics and Astronomy, University College London, London, WC1E 6B, UK\label{add5} \and School of Science \& Technology, City St. George’s, University of London, London, EC1V 0HB, UK\label{add4} \and
  Department of Physics and Astronomy, Stony Brook University, Stony Brook, NY 11794, USA\label{add2} \and
  Laboratoire de Physique Théorique et Hautes Énergies (LPTHE), UMR 7589, Sorbonne Université et CNRS, 4 place Jussieu, 75252 Paris Cedex 05, France\label{add1} \and
  Department of Physics, Miami University, 500 E. Spring St., Oxford, OH 45056, USA\label{add3}
}

\date{\today}
\maketitle

\begin{abstract}
  We present version 1.11 of \madanalysis, which extends the software package in several major ways to improve the handling of efficiency tables, the computation of observables in different reference frames and the calculation of statistical limits and/or significance. We detail how these improvements, whose development was motivated by the desire to implement two Run 2 LHC analyses targeting signatures with soft leptons and missing energy and exhibiting mild excesses (ATLAS-SUSY-2018-16 and ATLAS-SUSY-2019-09), have been implemented by both direct extensions of the code and integrations with third-party software. We then document the implementation and validation of these analyses, demonstrating their utility along with the improved statistics capabilities of \madanalysis through an investigation of the Next-to-Minimal Supersymmetric Standard Model in the context of a larger set of overlapping excesses in channels with soft leptons/jets and missing transverse energy.
\end{abstract}




\section{Introduction}

The LHC has conducted extensive searches for physics beyond the Standard Model (BSM), with results typically presented in terms of constraints on specific new physics scenarios. These constraints are often derived within simplified models, where the Standard Model (SM) is extended by a minimal number of fields and couplings, and they may not be directly applicable to more complex theoretical frameworks. To fully leverage the existing wealth of experimental data, the high-energy physics community has developed various strategies that enable the reinterpretation of existing LHC analyses in alternative models~\cite{LHCReinterpretationForum:2020xtr}. Many groups have also developed dedicated public software. 

These programs can be broadly classified into three categories. A first class of tools relies on simplified model results, where theoretical predictions for new physics signal cross sections (or event counts, incorporating selection efficiency information for a given signature) are compared with experimental upper bounds. This approach is computationally lighter and benefits from the simplicity of reducing complex models to a limited set of relevant signatures. However, it is inherently limited in scope, as it only constrains BSM scenarios that align well with the simplified model assumptions. Tools in this category include the \lstinline{SModelS}~\cite{Kraml:2013mwa, Ambrogi:2018ujg, Alguero:2020grj, Alguero:2021dig, Altakach:2024jwk} and \lstinline{DarKCast}~\cite{Ilten:2018crw} packages. A more precise and general approach is employed in a second class of tools that utilise Monte Carlo simulations to generate new physics signals emerging from a given model. This approach requires an accurate modelling of detector effects and the application of analysis selection criteria that reproduce the experimental analyses as closely as possible, which then allows for the computation of expected event yields. The obtained results can, in a second step, be statistically compared to observed data and SM predictions to derive constraints on the model. While this method is more robust and model-independent, it is computationally more demanding, often requiring significant resources. Public frameworks in this category of software include \madanalysis~\cite{Conte:2012fm, Conte:2014zja, Conte:2018vmg}, \lstinline{HackAnalysis}~\cite{Goodsell:2021iwc, Goodsell:2024aig}, \lstinline{ColliderBit}~\cite{GAMBIT:2017qxg} and \lstinline{CheckMate2}~\cite{Dercks:2016npn, Desai:2021jsa}. In the third category, an extensive number of LHC measurements (related to analyses of SM processes) have been systematically implemented within the \lstinline{Rivet} framework~\cite{Buckley:2010ar, Buckley:2019stt, Bierlich:2019rhm, Bierlich:2024vqo}. These measurements, originally designed to validate and refine predictions for SM processes, can also be leveraged to constrain BSM scenarios. The \lstinline{Contur} tool~\cite{Butterworth:2016sqg, Buckley:2021neu, CONTUR:2025yis} automates this process by assessing potential deviations from the SM in a wide range of parton-level differential distributions, providing a complementary approach to conventional BSM investigations. 

The \madanalysis\ package~\cite{Conte:2012fm, Conte:2014zja, Conte:2018vmg} was released more than a decade ago as a dedicated framework for collider phenomenology. It provides a user-friendly environment for designing new phenomenological studies and for recasting results from existing searches for new physics at colliders. The platform features two operational modes: a normal mode, which allows users to conduct phenomenological analyses without requiring advanced programming skills; and an expert mode, which offers full customisation and supports the implementation of existing LHC analyses. Once thoroughly validated, these implementations can be integrated into the code's Public Analysis Database (PAD)~\cite{Dumont:2014tja}, enabling systematic recasting of LHC results. Over the years, the expert mode, which serves as the foundation for LHC recasting, has undergone significant extensions incorporating several new functionalities. A key development has been the implementation of a built-in simplified fast detector simulation (SFS)~\cite{Araz:2020lnp}, which employs efficiency functions and smearing techniques to map the hadron-level Monte Carlo truth to reconstructed objects used in analyses. This framework is highly flexible, allowing efficiency functions to depend on a broad range of observables, including those relevant for long-lived particle searches~\cite{Araz:2021akd}. Recent progress also targets the integration of jet substructure techniques, particularly valuable for identifying highly boosted objects such as top quarks and weak bosons~\cite{Araz:2023axv}. 

Initially, the statistical interpretation of recasting results determined with \madanalysis\ was performed using computed yields and the CL$_s$ prescription~\cite{Read:2002hq}, applied independently to each signal region of an analysis and retaining the most sensitive one. The expected sensitivity was determined based on background estimates provided by the experimental collaborations, theoretical predictions for the signal under consideration (obtained by running the recast code), and observed event counts derived by incorporating experimental data. However, this approach was suboptimal as it did not account for correlations between disjoint signal regions. This was indeed a bottleneck in those days due to the lack of public statistical combination information. To address this limitation, the CMS and ATLAS collaborations began releasing simplified likelihoods~\cite{CMS-NOTE-2017-001} and full statistical models~\cite{ATL-PHYS-PUB-2019-029}, respectively, enabling a more rigorous statistical treatment by combining multiple signal regions. This development significantly improved the precision of exclusion limits obtained from recast codes, ensuring a more faithful exploitation of available experimental results. \madanalysis\ was upgraded to support these features a few years ago~\cite{Alguero:2022gwm}, now providing dedicated interfaces to incorporate the statistical material released by the collaborations. In addition, the limit-setting procedure as implemented in \madanalysis allows for straightforward extrapolations to higher luminosities, assuming constant signal and background efficiencies when extrapolating from one luminosity to another~\cite{Araz:2019otb}. 

Among the many searches for new physics, several analyses targeting simplified models inspired by the Minimal Supersymmetric Standard Model (MSSM) have reported excesses that have drawn significant attention from the high-energy physics community. For example, recent works~\cite{Canepa:2020ntc, Buanes:2022wgm, Cao:2022htd, Domingo:2022pde, Barman:2022jdg, Baum:2023inl, Stark:2023ont, Baer:2023olq, Cao:2023juc, Ashanujjaman:2023tlj, Carpenter:2023agq, Altakach:2023tsd, Baer:2023ech, Roy:2024yoh, Baer:2024kms, Chakraborti:2024pdn, Martin:2024pxx, Agin:2024yfs, Ellwanger:2024vvs, Capdevilla:2024bwt, Fuks:2024qdt, Martin:2024ytt, Agin:2025vgn} have examined several intriguing excesses emerging from searches for light electroweakinos decaying into missing transverse energy and soft objects. In~\cite{Agin:2023yoq, Agin:2024yfs, Fuks:2024qdt, Agin:2025vgn}, it was pointed out that some of these excesses potentially overlap when interpreted in terms of simplified supersymmetric scenarios in which the light electroweakinos consist predominantly of higgsinos or a bino-wino admixture with a compressed mass spectrum. In this case, these excesses arise in searches for new phenomena featuring missing transverse energy $\met$, in association with either jets~\cite{ATLAS:2021kxv, CMS:2021far} or soft leptons~\cite{ATLAS:2019lng, ATLAS:2021moa, CMS:2021edw}. Given the importance of scrutinising the excesses in these analyses, it is crucial for the community to have access to validated recast codes within public tools. This work contributes to that goal by introducing a major extension of \madanalysis, released as version 1.11, which introduces several new features to facilitate the implementation of the aforementioned searches. While monojet searches have been available for some time~\cite{CMS:2021far, Agin:2024yfs}, new methods were required to recast soft-lepton analyses~\cite{ATLAS:2019lng, ATLAS:2021moa}. The present work, therefore, enables the automatic confrontation of data with predictions from new physics models in the framework of currently interesting LHC excesses, using \madanalysis within a robust computational pipeline. Additionally, we enhance the code's limit-setting capabilities by interfacing it with the package \spey~\cite{Araz:2023bwx}, which provides a variety of statistical methods to assess whether a given signal is excluded.

The rest of this manuscript is organised as follows. Section~\ref{sec:ma5_new} details the novelties of \madanalysis version 1.11, which lie at the heart of this work. In section~\ref{sec:atlas_recasts}, we demonstrate their usage by implementing two ATLAS soft-lepton searches~\cite{ATLAS:2019lng, ATLAS:2021moa} in \madanalysis, providing detailed validations. Furthermore, in section~\ref{sec:application}, we illustrate how the set of validated LHC analysis implementations available within \madanalysis\ can be used to constrain new physics, taking the example of the Next-to-Minimal Supersymmetric Standard Model~\cite{Ellwanger:2009dp} (NMSSM) with a focus on the aforementioned excesses (and therefore the considered novelties of the code). Finally, we summarise our work and conclude in section~\ref{sec:conclusion}.


\section{Novelties of \texorpdfstring{\madanalysis}~version 1.11}\label{sec:ma5_new}
\madanalysis\ operates in two distinct running modes: the so-called normal and expert modes, each catering to different user needs and levels of expertise. The \textit{normal mode} is designed for users who wish to perform collider phenomenology studies without requiring advanced programming skills. It provides a straightforward, command-line-driven environment where users can analyse event samples, apply predefined selection cuts, reconstruct physical objects and generate histograms and cut-flows with minimal effort. In contrast, the \textit{expert mode} is more advanced and enables full customisation of the analysis pipeline. It is designed for users who need complete control over event selection, object reconstruction and the definition of selection cuts. It relies on a built-in C++ core, allowing direct access to the event record and enabling the implementation of sophisticated analysis strategies that closely mimic those used by the LHC experimental collaborations. Over the years, the expert mode has been continuously extended with new functionalities, such as a new detector simulation module~\cite{Araz:2020lnp, Araz:2021akd} and advanced jet substructure techniques~\cite{Araz:2023axv}, to meet the growing demands for efficiency and versatility in LHC recasting.

In this context, we introduce in section~\ref{sec:efficiencies} a new class, \lstinline{Efficiency1D}, along with associated methods in the internal C++ data structure of the code. This class has been specifically designed to handle one-dimensional efficiency functions as provided in \lstinline{HEPData}. In section~\ref{sec:lorentz}, we present updates to the C++ core of the \madanalysis framework related to Lorentz rotations and boosts, which are particularly useful for computing observables in different reference frames within an analysis. These methods play a key role in interfacing \madanalysis with the \lstinline{RestFrames} package~\cite{Jackson:2017gcy}, essential for the computation of the jigsaw variables~\cite{Jackson:2016mfb} required for the considered searches for supersymmetry with soft leptons. This interface is described in section~\ref{sec:restframes}. Finally, section~\ref{sec:spey} briefly describes updates in the limit setting procedure used in \madanalysis, which now entirely relies on the third-party software \spey~\cite{Araz:2023bwx} and its large set of built-in methods.

\subsection{One-dimensional efficiencies}\label{sec:efficiencies}
To facilitate the use of one-dimensional efficiency functions in \madanalysis, we introduce the \lstinline{Efficiency1D} class. This class provides a streamlined interface for handling efficiency lookup tables encoded as a function of a given kinematic variable such as the transverse momentum $p_T$ or the transverse energy $E_T$ of a reconstructed object. In the case of LHC recasting, these tables are typically extracted from external data sources such as \lstinline{HEPData} where the ATLAS and CMS collaborations officially publish efficiency maps encoded on a CSV format. The \lstinline{Efficiency1D} class then enables users to automatically read efficiency values from CSV files, initialising an instance of the class in this way, and retrieve them during runtime through interpolation based on the relevant input kinematic variable. 

The \lstinline{Efficiency1D} class assumes that input CSV files contain ordered rows with six entries, where each row is associated with a specific bin in the efficiency table. The first three entries of a row specify the central value, lower bound, and upper bound of a bin, while the last three entries correspond to the efficiency value and its upper and lower uncertainties (set to zero if not relevant). An efficiency table can be loaded upon instantiation of an \lstinline{Efficiency1D} object as follows:
\begin{lstlisting}[language=C++]  
  Efficiency1D *MuonEff = 
    new Efficiency1D(<std::string filename>);
\end{lstlisting}
Alternatively, the CSV file can be read separately using the \lstinline{ReadCSV} method:
\begin{lstlisting}[language=C++]  
  Efficiency1D *ElecEff = new Efficiency1D();
  ElecEff->ReadCSV(<std::string filename>);
\end{lstlisting}
Here, \lstinline{filename} is the path to the CSV file containing the efficiency data, encoded as a string. Once the table is loaded, efficiency values can be accessed using the \lstinline{Get} method of the \lstinline{Efficiency1D} class. For example, assuming an efficiency table depending on $p_T$, one can retrieve the efficiency for a muon with $p_T = 75$~GeV using:
\begin{lstlisting}[language=C++]
  MAdouble64 pt = 75.0;
  if(MuonEff->Initialised())
    MAdouble64 efficiency = MuonEff->Get(pt);
\end{lstlisting}
The \lstinline{Initialised} method ensures that the efficiency table has been successfully loaded before attempting to access values, and we remind that \lstinline{MAdouble64} is the internal \madanalysis type for double-precision floating-point numbers, ensuring portability across different platforms~\cite{Conte:2018vmg}.

\subsection{Rotations and boosts}\label{sec:lorentz}
\madanalysis provides various methods for rotating and boosting four-vectors between different reference frames. However, these methods have not been documented until now. Given their crucial role in the implementation of soft-lepton searches exhibiting excesses in which some combinations of four-momenta need to be transformed between frames, we provide a description of these transformations in this section. In previous versions of the code, rotations could be performed in two ways; a third manner was added from version 1.11 onward.

The \lstinline{MARotation3axis} class offers dedicated methods to rotate a four-vector by a specified angle around one of the three Cartesian axes defining the laboratory frame. The rotation angle and axis of the rotation can be specified either when instantiating an object of the class,
\begin{lstlisting}
  MARotation3axis *rot =  new MARotation3axis(
    <MAdouble64 angle>, <AxisType axis>
  );
\end{lstlisting}
or later by setting them explicitly,
\begin{lstlisting}
  MARotation3axis* rot;
  rot->setAngleAxis(
    <MAdouble64 angle>, <AxisType axis>
  );
\end{lstlisting}
Here, the second argument \lstinline{axis} can take one of three values \lstinline{Xaxis}, \lstinline{Yaxis}, or \lstinline{Zaxis}, which corresponds to rotations of an angle provided through the argument \lstinline{angle} around the $x$-axis, $y$-axis and $z$-axis, respectively. To apply the defined rotation to a four-vector $q^\mu$, the \lstinline{rotate} method of the class is used. This method takes an \lstinline{MALorentzVector} object as an argument, which is the internal \madanalysis format for handling four-vectors, and updates the value of the components of the input four-vector. The action on an illustrative four-vector $q^\mu = (1,2,3,4)$ could be implemented as
\begin{lstlisting}
  MALorentzVector q(1.,2.,3.,4.);
  rot->rotate(q);
\end{lstlisting}

More general rotations around arbitrary axes can be performed by specifying the three associated Euler angles $\varphi$, $\theta$ and $\psi$. These angles define a sequence of rotations that fully describe the orientation of a coordinate system relative to another. The spatial components of a four-vector $q^\mu = (q^0, \vec{q})$ are then transformed by multiplying them with the corresponding (passive) rotation matrix $\mathcal{M}$, following  $\vec{q} \to \mathcal{M} \vec{q}$. In the \madanalysis convention, this rotation matrix is related to the Euler angles according to
\be\setlength\arraycolsep{4pt}
  \mathcal{M} = \bpm
    c_\varphi c_\psi - s_\varphi s_\psi c_\theta &  s_\varphi c_\psi + c_\varphi s_\psi c_\theta & s_\psi s_\theta\\
    -c_\varphi s_\psi - s_\varphi c_\psi c_\theta &  -s_\varphi s_\psi + c_\varphi c_\psi c_\theta & c_\psi s_\theta\\
    s_\theta s_\varphi & -s_\theta c_\varphi & c_\theta
  \epm\,,
\ee
where $c_x$ and $s_x$ denote the cosine and sine of the angle $x$, respectively. To apply such rotations, the user must instantiate an object of the \lstinline{MARotation3euler} class. This can be done in a single step,
\begin{lstlisting}
  MARotation3euler *rot = new MARotation3euler(
    <MAdouble64 phi>,
    <MAdouble64 theta>, 
    <MAdouble64 psi>
  );
\end{lstlisting}  
or in two steps by setting the angles explicitly,
\begin{lstlisting}
  MARotation3euler* rot;
  rot->setAngles(
    <MAdouble64 phi>,
    <MAdouble64 theta>,
    <MAdouble64 psi>
  );
\end{lstlisting}  
Here, the three Euler angles $\varphi$, $\theta$ and $\psi$ are passed as the double-precision arguments \lstinline{phi}, \lstinline{theta} and \lstinline{psi}. Once the rotation object is configured, the transformation of a four-vector $q^\mu$ proceeds in the same way as for the \lstinline{MARotation3axis} class. The user simply calls the \lstinline{rotate} method equipping the class, and applies it to a \lstinline{MALorentzVector} object. For the same four-vector $q^\mu = (1, 2, 3, 4)$ as the one introduced before,  this would explicitly give
\begin{lstlisting}
  MALorentzVector q(1.,2.,3.,4.);
  rot->rotate(q);
\end{lstlisting}  

In some cases, it is more convenient to define a rotation using a single angle and a specific axis around which vectors are rotated, rather than being limited to the $x$-axis, $y$-axis or $z$-axis of the laboratory frame or to the three Euler angles. The \madanalysis framework introduces a new class, \lstinline{MARotationGeneral}, in version 1.11 to allow for such general rotations. This feature provides additional flexibility when transforming four-vectors in non-standard reference frames, which is, in particular, relevant for the integration of the \lstinline{RestFrames} package~\cite{Jackson:2017gcy} discussed in section~\ref{sec:restframes}. As with previous rotation classes, a general rotation can be implemented in either one or two steps. The user may either instantiate the \lstinline{MARotationGeneral} object directly with the rotation angle and axis,  
\begin{lstlisting}
  MARotationGeneral *rot=new MARotationGeneral(
    <MAdouble64 angle>,
    <MAVector3 dir>
  );
\end{lstlisting}  
or set the parameters separately after object creation,  
\begin{lstlisting}
  MARotationGeneral* rot;
  rot->setAngleAxis(
    <MAdouble64 angle>,
    <MAVector3 dir>
  );
\end{lstlisting}  
In both cases, the rotation angle is the double-precision number \lstinline{angle} and the axis of rotation is specified as a three-component vector (encoded as a \lstinline{MAVector3} object in the \madanalysis data format). This vector does not need to be normalised beforehand, as normalisation is handled internally by the code. Denoting the rotation angle by $\varphi$ and the (normalised) axis direction by $\vec{e}$, the spatial components $\vec{q}$ of a four-vector $q^\mu = (q^0, \vec{q})$ are transformed according to Rodrigues’ rotation formula,
\begin{equation}
  \vec{q} \to c_\varphi \vec{q} \ +\  (\vec{q} \times \vec{e}) s_\varphi \ +\ (1 - c_\varphi) (\vec{q} \cdot \vec{e}) \vec{e}\,.
\end{equation}
which can be used in the code by calling the method \lstinline{rotate} of the class with a four-vector given in argument,
\begin{lstlisting}
    rot->rotate(<MALorentzVector q>);
\end{lstlisting}
An example illustrating the usage of this class could simply be:
\begin{lstlisting}
  MAVector3 axis(1., 1., 1.);
  MAdouble64 phi = M_PI/3.;
  MARotationGeneral* rot=new MARotationGeneral(
    phi, axis
  );
  MALorentzVector q(1.,2.,3.,4.);
  rot->rotate(q);
\end{lstlisting} 

Boosts can be implemented in a similar manner using the \lstinline{MABoost} class and its associated methods. Given a velocity vector $\vec{\beta} = (\beta_x, \beta_y, \beta_z)$, a boost object can be instantiated directly in a single step,
\begin{lstlisting}
  MABoost *booster =new MABoost(
    <MAdouble64 bx>,
    <MAdouble64 by>,
    <MAdouble64 bz>
  );
\end{lstlisting}  
Alternatively, the boost parameters can also be set separately after declaring an \lstinline{MABoost} object:  
\begin{lstlisting}
  MABoost *booster;
  booster->setBoostVector(
    <MAdouble64 bx>,
    <MAdouble64 by>,
    <MAdouble64 bz>
  );
\end{lstlisting}
with the three components $\beta_x$, $\beta_y$ and $\beta_z$ of the boost vector being passed as the double-precision arguments \lstinline{bx}, \lstinline{by} and \lstinline{bz}. Applying a boost with velocity $\vec{\beta}$ to a four-vector $q^\mu = (q^0, \vec{q})$ transforms its components according to
\be\setlength\arraycolsep{8pt}
  q^\mu \to \bpm 
    \gamma \big(q^0+\vec{\beta}\cdot\vec{q}\big)\\
    \vec{q} + \bigg( \frac{\gamma^2}{1+\gamma}\, (\vec{\beta}\cdot\vec{q})+ \gamma q^0 \bigg) \vec{\beta}
  \epm\,,
\ee
where $\gamma = 1/\sqrt{1-|\vec{\beta}|^2}$ is the Lorentz factor. This transformation is implemented using the \lstinline{boost} method of the \lstinline{MABoost} class, which takes a four-vector as an argument (encoded as an \lstinline{MALorentzVector} object),
\begin{lstlisting}
    booster->boost(<MALorentzVector q>);
\end{lstlisting}
A simple example illustrating how to apply a boost in a \madanalysis-based analysis implementation could be given by:
\begin{lstlisting}
  MAVector3 velocity(0.1, 0.2, 0.3);
  MABoost *booster = new MABoost(
    velocity.X(),
    velocity.Y(),
    velocity.Z()
  );
  MALorentzVector q(1.,2.,3.,4.);
  booster->boost(q);
\end{lstlisting} 

\subsection{Integration of the \texorpdfstring{\lstinline{RestFrames}}~ package}\label{sec:restframes}
Recursive jigsaw reconstruction is a methodology developed to resolve combinatorial and kinematic ambiguities in the reconstruction of events involving the pair production of new particles, followed by their cascade decays into both visible and invisible final state objects. This technique systematically identifies the minimal set of unknowns required to determine complementary observables built from event-level quantities in the laboratory frame as well as in successive reference frames associated with the decaying parent particles.

The \lstinline{RestFrames} library was initially released as a stand-alone package relying on \lstinline{Root} for various functionalities, including vector and matrix manipulations as well as minimisation routines. While \madanalysis itself is interfaced with \lstinline{Root}, we designed its \lstinline{RestFrames} integration in such a way that it remains functional even when \lstinline{Root} is unavailable on the system. In order to interface \lstinline{RestFrames} with \madanalysis, all references to \lstinline{Root} in its source code had to be replaced with calls to methods inherent to \madanalysis, including the boost and rotation functions described in section~\ref{sec:lorentz}. To achieve this, we followed the strategy outlined in~\cite{Agin:2024yfs} and employed in \lstinline{HackAnalysis} version 2. In practice, all references to three-component vectors and four-vectors now rely on the internal data format of \madanalysis, minimisation routines utilise the Nelder-Mead algorithm~\cite{Nelder:1965zz} that has been incorporated into the code, and matrix operations are performed using the \lstinline{Eigen} package, available at \href{https://gitlab.com/libeigen/eigen}{https://gitlab.com/libeigen/eigen}.

The \lstinline{PHYSICS} service of \madanalysis has been extended to include a helper pointer, \lstinline{RF}, which provides direct access to \lstinline{RestFrames} methods. In recursive jigsaw reconstruction, events are structured into a so-called decay tree, designed to reflect the kinematic features characteristic of the studied signal topology. For instance, in analyses relevant to the electroweakino excesses at the heart of this work, visible objects are grouped into two distinct visible frames: one associated with the initially produced system of (decaying) supersymmetric particles (\lstinline{S}) and another related to initial-state radiation (\lstinline{ISR}) which remains non-decaying. These frames are used to reconstruct the centre-of-mass frame (\lstinline{CM}), which is itself related to the laboratory frame (\lstinline{LAB}). The setup of these frames can be initialised within \madanalysis via the dedicated \lstinline{PHYSICS->RF} helper function,
\begin{lstlisting}
  PHYSICS->RF->addLabFrame("LAB");
  PHYSICS->RF->addDecayFrame("CM");
  PHYSICS->RF->addDecayFrame("S");
  PHYSICS->RF->addVisibleFrame("ISR");
\end{lstlisting}
Additional frames can be introduced in a similar manner. For instance, the supersymmetric system decays into both visible (\lstinline{V}) and invisible (\lstinline{I}) objects. These can be associated with dedicated frames as
\begin{lstlisting}
  PHYSICS->RF->addVisibleFrame("V");
  PHYSICS->RF->addInvisibleFrame("I");
\end{lstlisting}
Each of the above functions serves a distinct purpose. The method \lstinline{addLabFrame(<std::string name>)}  defines the laboratory frame, which serves as the reference frame for all event-level quantities. Next, the method \lstinline{addDecayFrame(<std::string name>)} introduces a frame corresponding to a decaying particle or system of particles, allowing for the stepwise reconstruction of event kinematics. Finally, the complementary methods \lstinline{addVisibleFrame(<std::string name>)} and  \lstinline{addInvisibleFrame(<std::string name>)} define frames associated with visible and invisible final-state particles, which are directly and indirectly reconstructed from detector-level objects, respectively. These functions allow for the flexible construction of decay trees tailored to the specific kinematics of the studied signal. The connection of the leaves of the tree (\ie\ the different introduced frames) is achieved via conventional methods of the \lstinline{RestFrames} package, like \lstinline{SetChildFrame} and \lstinline{AddChildFrame}. We refer to the documentation of the package for examples, as well as to the implementation code of the ATLAS-SUSY-2018-16 search which is available from the \madanalysis dataverse\footnote{\href{https://doi.org/10.14428/DVN/1CPXHX}{https://doi.org/10.14428/DVN/1CPXHX}}~\cite{1CPXHX_2025} and which is further detailed below as a concrete illustration. Here, the \lstinline{RestFrames} decay tree targets a topology in which a compressed electroweakino pair is produced with an initial-state-radiation jet. The centre-of-mass frame (\lstinline{CM}) thus decays into an \lstinline{ISR} system and a supersymmetric decay system \lstinline{S}, the latter collecting visible leptons \lstinline{L}, other soft objects \lstinline{V} and invisible particles \lstinline{I}. This decay tree is correspondingly initialised as
\begin{lstlisting}
  PHYSICS->RF->addLabFrame("LAB");
  PHYSICS->RF->addDecayFrame("CM");
  PHYSICS->RF->addDecayFrame("S");
  PHYSICS->RF->addVisibleFrame("ISR");
  PHYSICS->RF->addVisibleFrame("V");
  PHYSICS->RF->addVisibleFrame("L");
  PHYSICS->RF->addInvisibleFrame("I");
\end{lstlisting}
where \lstinline{LAB} refers to the laboratory frame. The tree topology is further defined through
\begin{lstlisting}
  PHYSICS->RF->getLabFrame("LAB")->SetChildFrame(*(PHYSICS->RF->getDecayFrame("CM")));
  PHYSICS->RF->getDecayFrame("CM")->AddChildFrame(*(PHYSICS->RF->getVisibleFrame("ISR")));
  PHYSICS->RF->getDecayFrame("CM")->AddChildFrame(*(PHYSICS->RF->getDecayFrame("S")));
  PHYSICS->RF->getDecayFrame("S")->AddChildFrame(*(PHYSICS->RF->getVisibleFrame("V")));
  PHYSICS->RF->getDecayFrame("S")->AddChildFrame(*(PHYSICS->RF->getVisibleFrame("L")));
  PHYSICS->RF->getDecayFrame("S")->AddChildFrame(*(PHYSICS->RF->getInvisibleFrame("I")));
\end{lstlisting}
encoding the decay chains \lstinline{LAB} $\to$ \lstinline{CM} $\to$ \{\lstinline{ISR}, \lstinline{S}\} and \lstinline{S} $\to$ \{\lstinline{V}, \lstinline{L}, \lstinline{I}\}.

Jigsaw reconstruction strongly relies on groups of objects whose four-momenta can be combined to reconstruct the four-momentum of a parent particle or to define specific observables like estimators for the ratio of the missing transverse momentum $\met$ to the $p_T$ of an initial state radiation, named $R_\mathrm{ISR}$ in the jigsaw reconstruction framework. Two main types of groups can be defined: invisible groups and combinatorial groups. The former, implemented via the function
\begin{lstlisting}
  PHYSICS->RF->addInvisibleGroup(
    <std::string name>
  );
\end{lstlisting}
includes invisible objects in an event. The latter, initialised with
\begin{lstlisting}
  PHYSICS->RF->addCombinatoricGroup(
    <std::string name>
  );
\end{lstlisting}
gathers visible objects with the purpose of further splitting them into subgroups according to specific jigsaw rules. In the ATLAS-SUSY-2018-16 example, invisible and visible objects are assigned to two groups which thus interface between the abstract decay tree and the reconstructed objects,
\begin{lstlisting}
  PHYSICS->RF->addInvisibleGroup("INV");
  PHYSICS->RF->addCombinatoricGroup("VIS");
\end{lstlisting}
The \lstinline{INV} group collects all invisible frames (here only \lstinline{I}) and is responsible for reconstructing the invisible four-momentum from the missing transverse energy,
\begin{lstlisting}
 PHYSICS->RF->getInvisibleGroup("INV")->AddFrame(*(PHYSICS->RF->getInvisibleFrame("I")));
\end{lstlisting}
The \lstinline{VIS} group is instead a combinatoric group, meaning that it receives a collection of reconstructed visible objects dynamically assigned to the visible frames of the tree (\lstinline{ISR} and \lstinline{V}), with constraints on the minimum number of objects per frame,
\begin{lstlisting}
  PHYSICS->RF->getCombinatoricGroup("VIS")->AddFrame(*(PHYSICS->RF->getVisibleFrame("ISR")));
  PHYSICS->RF->getCombinatoricGroup("VIS")->SetNElementsForFrame(*(PHYSICS->RF->getVisibleFrame("ISR")), 1, false);
  PHYSICS->RF->getCombinatoricGroup("VIS")->AddFrame(*(PHYSICS->RF->getVisibleFrame("V")));
  PHYSICS->RF->getCombinatoricGroup("VIS")->SetNElementsForFrame(*(PHYSICS->RF->getVisibleFrame("V")), 0, false);
\end{lstlisting}
The second argument to \lstinline{SetNElementsForFrame} specifies the minimum number of objects assigned to that frame, while the third argument (\lstinline{false}) indicates that this number is not fixed, \ie\ additional objects may be assigned to it. Here, \lstinline{ISR} is required to receive at least one object (ensuring thus a valid \lstinline{ISR} hemisphere), while \lstinline{V} may receive zero or more. On the other hand, the \lstinline{L} frame which collects the soft leptons is not added to the combinatoric group as leptons are identified unambiguously.

Several jigsaw rules are implemented in \lstinline{RestFrames} to determine the properties of parent or invisible particles. In the case of invisible particles, the various states can be identified using contra-boost invariant, invisible mass and invisible rapidity jigsaw rules, all defined in~\cite{Jackson:2017gcy}. For visible objects, subgroup separation as required, for instance, to isolate the two decay chains arising from the pair production of parent supersymmetric particles, can be achieved by minimising specific dimensionful mass parameters. The corresponding jigsaw rules available in \lstinline{RestFrames} can be set, in the \madanalysis framework, through functions like
\begin{lstlisting}
  PHYSICS->RF->addInvisibleJigsaw(
    <std::string name>, 
    RestFramesHelper::<keyword>
  );
  
  PHYSICS->RF->addCombinatoricJigsaw(
    <std::string name>,
    RestFramesHelper::<keyword>
  );
\end{lstlisting}
where the first argument specifies the name of the jigsaw rule, and the second argument refers to the method governing the rule. The currently implemented methods include \lstinline{kContraBoost}, \lstinline{kSetMass} and \lstinline{kSetRapidity}, for invisible jigsaw rules relying on contra-boost invariants, invisible mass and invisible rapidity, respectively, while \lstinline{kMinMasses} is available for combinatoric jigsaw rules built on some reconstructed mass quantity. Once defined, these groups can be manipulated using all relevant methods provided by the \lstinline{RestFrames} package, ultimately allowing for the computation of key observables to distinguish a supersymmetric signal from the corresponding SM background. A complete list of available methods is documented in the \lstinline{RestFrames} manual and code. In a \madanalysis implementation, access to the various objects involved in jigsaw reconstruction and initialised in the code is facilitated by dedicated accessors to be called from the \madanalysis helper \lstinline{PHYSICS->RF}, which takes a frame name as an argument. They include accessors to frames of reference comprising \lstinline{getLabFrame}, \lstinline{getVisibleFrame}, \lstinline{getDecayFrame} and \lstinline{getInvisibleFrame}, as well as accessors to jigsaw rules like \lstinline{getInvisibleJigsaw} or \lstinline{getCombinatoricJigsaw}, and accessors to the two classes of groups that could be defined in an analysis implementation, \lstinline{getCombinatoricGroup} and \lstinline{getInvisibleGroup}.

In the ATLAS-SUSY-2018-16 analysis, two jigsaw rules resolve the unknowns of the system: an \lstinline{InvMass} jigsaw that sets the invariant mass of the invisible system and a \lstinline{SplitVis} combinatoric jigsaw that partitions the reconstructed objects between the \lstinline{ISR} and \lstinline{S} hemispheres by minimising the sum of their masses,
\begin{lstlisting}
 PHYSICS->RF->addInvisibleJigsaw("InvMass", RestFramesHelper::kSetMass);
 PHYSICS->RF->addCombinatoricJigsaw("SplitVis", RestFramesHelper::kMinMasses);
 PHYSICS->RF->getInvisibleGroup("INV")->AddJigsaw(*(PHYSICS->RF->getInvisibleJigsaw("InvMass")));
 PHYSICS->RF->getCombinatoricGroup("VIS")->AddJigsaw(*(PHYSICS->RF->getCombinatoricJigsaw("SplitVis")));
\end{lstlisting}
The \lstinline{SplitVis} jigsaw is further instructed which frames belong to which hemisphere so that it knows how to partition the objects,
\begin{lstlisting}
 PHYSICS->RF->getCombinatoricJigsaw("SplitVis")->AddFrame(*(PHYSICS->RF->getVisibleFrame("ISR")), 0);
 PHYSICS->RF->getCombinatoricJigsaw("SplitVis")->AddFrame(*(PHYSICS->RF->getVisibleFrame("V")), 1);
 PHYSICS->RF->getCombinatoricJigsaw("SplitVis")->AddFrame(*(PHYSICS->RF->getInvisibleFrame("I")), 1);
 PHYSICS->RF->getCombinatoricJigsaw("SplitVis")->AddFrame(*(PHYSICS->RF->getVisibleFrame("L")), 1);
\end{lstlisting}
The integer labels identify the two hemispheres: \lstinline{0} corresponds to the initial-state-radiation side and \lstinline{1} to the supersymmetric system. By assigning \lstinline{V}, \lstinline{I} and \lstinline{L} to hemisphere \lstinline{1}, the jigsaw is instructed to minimise the sum of masses of the two sides while respecting this hemisphere assignment, thus optimally separating the initial-state radiation from the supersymmetric system.

Once \lstinline{InitializeAnalysis()} is called, all \lstinline{RestFrames} kinematic variables become available for use in the analysis. After resetting the previous state with \lstinline{ClearEvent()}, jets are added to the combinatoric group \lstinline{VIS} one by one via \lstinline{AddLabFrameFourVector()}, with their pseudo-rapidity set to zero to work in the transverse plane. Leptons are instead summed into a single four-vector and assigned directly to the \lstinline{L} frame via \lstinline{SetLabFrameFourVector()}, reflecting the fact that their hemisphere assignment is unambiguous. The missing transverse momentum is finally passed to the invisible group \lstinline{INV} as a transverse three-vector. Calling \lstinline{AnalyzeEvent()} executes the jigsaw rules and allows for the computation of any key observables. For example, the $R_\mathrm{ISR}$ observable can be implemented as
\begin{lstlisting}
 MAVector3 P_ISR = PHYSICS->RF->getVisibleFrame("ISR")->GetFourVector(*(PHYSICS->RF->getDecayFrame("CM"))).Vect();
 MAVector3 P_INV = PHYSICS->RF->getInvisibleFrame("I")->GetFourVector(*(PHYSICS->RF->getDecayFrame("CM"))).Vect();
 MAdouble64 RISR = fabs(P_INV.Dot(P_ISR.Unit()) / P_ISR.Mag());
\end{lstlisting}

\subsection{Limit setting with \texorpdfstring{\lstinline{spey}}~}
\label{sec:spey}
Previously, limit setting in \madanalysis was performed by generating a large number of toy experiments in which the expected number of background events populating the signal regions of a given analysis was generated according to a Gaussian distribution, with mean and width extracted from information provided in the associated experimental publication. This background event count, combined with the predicted number of signal events in each signal region, was then used to generate the actual number of background and signal events according to Poisson distributions. From these, background and signal+background $p$-values were subsequently determined, and an exclusion limit confidence level inferred~\cite{Read:2002hq}. This simple procedure was generalised~\cite{Alguero:2022gwm} when correlation information between different signal regions became available, either in the form of a covariance matrix~\cite{CMS-NOTE-2017-001} or through the full statistical model used in the analysis~\cite{ATL-PHYS-PUB-2019-029} (or a simplified version of it). In the updated version of \madanalysis presented in this manuscript, we improve upon this procedure by utilising the third-party Python package \lstinline{spey}~\cite{Araz:2023bwx, spey_zenodo}, which enables the use of different likelihood prescriptions for limit setting. This functionality subsequently requires the installation of the Python packages \lstinline{spey} and \lstinline{spey_pyhf}~\cite{spey_pyhf_zenodo}.

By default, \madanalysis now uses \lstinline{spey} in three distinct ways, in a manner that is fully transparent and automatic for the user. If no background correlation information is available in the analysis implementation, the likelihood distribution for each region defined in the analysis is independently constructed using a Poisson distribution for the signal region counts and a unit Gaussian for the background uncertainties provided by the experimental collaboration. Observed and expected exclusion confidence limits are then computed for each region, and the most sensitive region is indicated in the output as the best, following the standard convention of defining it from the expected limits in order to avoid any bias from the observed data.

If correlations between regions are instead provided in the form of a covariance matrix, a combined or simplified likelihood distribution is built using Poisson distributions for all regions and a unit multivariate Gaussian to model uncertainties and correlations. In the case of multiple sets of correlated regions, the most sensitive one is again displayed as the best outcome.

Finally, if a \lstinline{HistFactory} model~\cite{ATL-PHYS-PUB-2019-029} is available, \lstinline{spey} directly employs it via its \lstinline{pyhf} plugin to construct the likelihood. As before, if more than one \lstinline{HistFactory} model is provided, the most sensitive one is selected and reported.

By default, the code computes both observed and \textit{a posteriori} exclusions (\lstinline{spey.ExpectationType.observed} and \lstinline{spey.ExpectationType.aposteriori} (post-fit) in the \lstinline{spey} language\footnote{See the \href{https://spey.readthedocs.io}{\lstinline{spey} documentation} for details.}). The former measures how extreme the observed result is compared to the background-only hypothesis, while the latter represents the typical $p$-value expected in repeated background-only experiments where the nuisance parameters are fitted to the observed data. Consequently, it serves as a benchmark for assessing the experiment's sensitivity. This contrasts with earlier implementations in \madanalysis where expected limits were derived using only pre-experiment information, without incorporating the effect of observed data. Such an approach would instead correspond, in the \spey\ terminology, to \lstinline{spey.ExpectationType.apriori} (pre-fit). It is now only used to extrapolate the results to luminosities different from the nominal value, following the guidelines in~\cite{Araz:2019otb}. In this case, the expected exclusions are thus computed by fitting the expected background yields under the assumption that the SM is the true model. The new limit-setting strategy included in \madanalysis therefore ensures consistency with the standard methodology used for sensitivity projections and aligns well with the results reported by experimental collaborations. 

This new approach to handling the statistical treatment of a recast run in \madanalysis\ by means of \spey renders the \lstinline{CLs_numofexps} attribute of the \lstinline{main.recast} class obsolete. Previously, this attribute specified the number of toy experiments conducted when computing CL$_\mathrm{s}$ exclusions, as detailed in~\cite{Dumont:2014tja,Conte:2018vmg}, but it is no longer relevant for exclusions computed within \spey as the computation has been standardised through the usage of asymptotic formulae. Instead, a new option is available to allow users to disable the statistical treatment and compute only the signal-region efficiencies of the selected analyses. This is achieved by setting, within the command line interpreter of \madanalysis,
\begin{lstlisting}
  set main.recast.analysis_only_mode = <bool>
\end{lstlisting}
where \lstinline{<bool>} is either \lstinline{True} or \lstinline{False} (default). When set to \lstinline{True}, the program will restrict itself to efficiency calculations without performing any limit-setting calculations. If this attribute is not specified (\ie\ it is set to \lstinline{False}, its default value), \madanalysis\ will run normally, computing both efficiencies and exclusion limits for the recast run considered.


\section{New ATLAS electroweakino searches in \texorpdfstring{\madanalysis}~}\label{sec:atlas_recasts}
In order to exemplify the usage of the new methods implemented in the \madanalysis data format introduced in section~\ref{sec:ma5_new}, we consider the implementation of the ATLAS-SUSY-2018-16 (see section~\ref{sec:atlas_susy_2018_16}) and the ATLAS-SUSY-2019-09 (see section~\ref{sec:atlas_susy_2019_09}) searches for compressed electroweakino signals. The choice of these searches is motivated by the fact that they have recently shown concurrent excesses of data with respect to the SM expectation.

\subsection{The ATLAS-SUSY-2018-16 search with soft dileptons}\label{sec:atlas_susy_2018_16}
\begin{table*}
\centering\renewcommand{\arraystretch}{1.4}\setlength{\tabcolsep}{12pt}
\begin{tabular}{l | l l}
  Variable  & $2\ell$  & $1\ell 1 T$\\
  \hline Number of leptons/tracks  & 2 leptons & 1~lepton~and~$\geq1$~track\\
  Lepton $p_T$~[GeV]  & $p_T(\ell_1)>5$ & $p_{T}(\ell)<10$\\
  $\Delta R_{\ell\ell}$ & $\Delta R_{ee}>0.30$, $\Delta R_{\mu\mu}>0.05$, $\Delta R_{e\mu}>0.2$ & $0.05<\Delta R_{\ell,\mathrm{track}}<1.5$\\
  Lepton charge and flavour & $e^{\pm}e^{\mp}$ or $\mu^{\pm}\mu^{\mp}$ & $e^{\pm}e^{\mp}$ or $\mu^{\pm}\mu^{\mp}$ \\
  \multirow{2}{*}{Di-lepton invariant mass~[GeV]} & $3<m_{ee}<60$, $1<m_{\mu\mu}<60$ & $0.5<m_{\ell,\mathrm{track}}<5$\\
  & Veto $3<m_{\ell\ell}<3.2$ & Veto $3<m_{\ell,\mathrm{track}}<3.2$\\
  $m_{\tau\tau}$~[GeV] & $<0$ or $>160$ & $-$ \\
  $\met$~[GeV] & $>120$ & $>120$ \\
  Number of jets & $\geq 1$ & $\geq 1$ \\
  Number of $b$-tagged jets & $=0$ & $-$ \\
  Leading jet $p_T$~[GeV] & $\geq 100$ & $\geq 100$\\
  $\min(\Delta\phi(\mathrm{any\,jet}, \ptmiss))$ & $>0.4$ & $>0.4$ \\
  $\Delta \phi(j_1,\vec{p}_T^{\text{miss}})$ & $\geq 2.0$ & $\geq 2.0$ \\
\end{tabular}\vspace{.2cm}\setlength{\tabcolsep}{9pt}
\begin{tabular}{l | l l l l}
  Variable &SR$-$E$-$low & SR$-$E$-$med &SR$-$E$-$high &SR$-$E$-1\ell 1 T$ \\
  \hline
  $\met$~[GeV] &$[120, 200]$ &$[120, 200]$ &$>200$ & $>200$ \\
  $\met/H_T^{\text{lep}}$ &$<10$ &$>10$ & $-$ & $>30$ \\
  $\Delta \phi(\ell, \ptmiss)$ & $-$ & $-$ & $-$ & $<1.0$ \\
  Lepton or track $p_T$~[GeV]\ \ \ \ \       & $p_T(\ell_2) > 5 + m_{\ell\ell}/4$ & $-$ &$p_T(\ell_2)>\min(10, 2 + m_{\ell\ell}/3)$ & $p_T(\text{track})<5$ \\
  $M_T^S$~[GeV] & $-$ &$<50$ & $-$ &$-$ \\
  $m_T(\ell_1)$~[GeV] &$[10,60]$ & $-$ &$<60$ &$-$ \\
  $R_{\text{ISR}}$ &$[0.8,1.0]$ & $-$ &$[\max(0.85,0.98-0.02\times m_{\ell\ell}),~1.0]$ & $-$ \\
\end{tabular}
\caption{Preselection (upper table) and selection (lower table) requirements defining the electroweakino search regions (SR$-$E) of the ATLAS-SUSY-2018-16 analysis.} \label{tab:201816_selection}
\end{table*}

The ATLAS-SUSY-2018-16 analysis~\cite{ATLAS:2019lng} searches for signals arising from the production and decay of electroweakino pairs in scenarios with compressed mass spectra. The analysis uses proton-proton collision data from the LHC Run~2 at $\sqrt{s} = 13$~TeV, corresponding to an integrated luminosity of 139~fb$^{-1}$, and targets events with low-momentum opposite-sign same-flavour (OSSF) leptons and missing transverse energy $\met$. Such a final state is expected to emerge from supersymmetric scenarios where the mass difference between the produced heavier supersymmetric particles and the lightest supersymmetric particle is small, leading to soft visible decay products. To enhance sensitivity, the analysis explores three event topologies, each associated with a specific set of signal regions (SRs). The first of these exploits the presence of initial-state radiation to boost the supersymmetric system (the so-called ‘electroweakino’ signal regions), while the other two focus on vector-boson fusion signatures and slepton pair production, respectively.

Our implementation targets the electroweakino SRs, labelled `SR-E’ in the ATLAS publication, which are further divided into two categories based on the multiplicity of reconstructed leptons. In the $2\ell$ channel, the final state comprises an OSSF lepton pair, whereas in the $1\ell1T$ channel, only a single lepton is required while the second is an isolated OSSF low-$p_T$ track. These two leptons (thus in the general sense) are ranked by transverse momentum, with $\ell_1$ and $\ell_2$ respectively denoting the leading and sub-leading leptons. For both categories of SRs, the analysis strategy exploits the presence of a high-momentum jet recoiling against the supersymmetric system, thereby increasing the missing transverse energy. 

Events are preselected based on missing momentum, lepton properties and hadronic activity to suppress SM backgrounds while preserving signal sensitivity. The upper panel in table~\ref{tab:201816_selection} details these preselection cuts. In particular, the final state should include at least one jet with $p_T\geq 100$~GeV and missing transverse energy of at least 120~GeV. Additionally, the missing transverse energy must be well separated from any reconstructed jet, especially the leading jet $j_1$, and the preselection also exploits the $m_{\tau\tau}$ variable~\cite{Han:2014kaa, Baer:2014kya, Barr:2015eva}, an estimator for the invariant mass of a pair of leptonically decaying tau leptons, to suppress the $Z\to\tau\tau$ background.

The selection requirements divide the $2\ell$ category into three classes of signal regions that are labelled `SR$-$E$-$high’, `SR$-$E$-$med' and `SR$-$E$-$low'. They differ based on the constraints imposed on the missing transverse momentum and the properties of the two leptons, as summarised in the lower panel of table~\ref{tab:201816_selection}. Specifically, different thresholds are applied to the missing transverse energy $\met$, its ratio to the leptonic activity $H_T^\mathrm{lep}$ (defined as the sum of the transverse momenta of the two leptons), the transverse momentum of the sub-leading lepton, and the transverse mass of the system comprising the leading lepton and the missing momentum.  Additionally, the selection includes constraints based on observables derived from the recursive jigsaw reconstruction technique. This method partitions the event final state into two hemispheres perpendicular to the thrust axis along which the electroweakinos recoil against initial-state radiation. The first hemisphere, referred to as the supersymmetric hemisphere, encompasses the entire supersymmetric system, including the decay products of the produced electroweakinos. The second hemisphere comprises instead the hadronic activity in the event. The definition of the signal regions then leverages two key observables, the transverse mass $M_T^S$ of the supersymmetric system and the ratio $R_\mathrm{ISR}$ of the missing transverse energy to the transverse momentum of the hadronic system. In contrast, the $1\ell1T$ selection, labelled `SR$-$E$-1\ell 1 T$', does not apply these additional cuts but instead requires the missing transverse momentum to be well separated from the leading lepton, as defined in the lower panel of table~\ref{tab:201816_selection}.

After applying the signal region-dependent selection cuts, the di-lepton invariant mass is binned such that each bin is taken to be an independent signal region. For the $2\ell$ regions, the invariant mass $m_{\ell\ell}$ is binned between 1~GeV and 60~GeV, while for the $1\ell1T$ regions, the invariant mass $m_{\ell,\text{track}}$ is binned between 0.5~GeV and 5.0~GeV. Additionally, in the $2\ell$ channel, the analysis distinguishes between different lepton-flavour combinations. While the original ATLAS analysis also includes an inclusive selection that merges all $2\ell$ and $1\ell1T$ bins into broader bins defined only by an upper bound on the di-lepton invariant mass, our recast does not include this inclusive approach. Instead, we rely exclusively on the more granular signal region definitions described above and make use of the \lstinline{pyhf} statistical model provided by the ATLAS collaboration for a statistical interpretation combining all these exclusive SRs.

As detailed in~\cite{Agin:2024yfs}, implementing the ATLAS-SUSY-2018-16 analysis within a recasting framework and validating the results presents several challenges. Thanks to the new version of \madanalysis introduced in this report, these challenges can now be addressed quite straightforwardly. One key aspect is the proper computation of observables based on the recursive jigsaw reconstruction technique, which is required to accurately reproduce ATLAS exclusion contours for specific benchmark scenarios, using the \lstinline{RestFrames} package. This is now feasible thanks to the new developments described in section~\ref{sec:restframes}. Additionally, the analysis relies on lepton reconstruction efficiencies extracted from figures available on \lstinline{HEPData}, and an event re-weighting procedure needs to be employed to account for missing transverse energy trigger efficiencies (also available on \lstinline{HEPData}). Such efficiencies can now be easily handled directly within a \madanalysis analysis implementation using the one-dimensional efficiency class introduced in section~\ref{sec:efficiencies}. 

\begin{figure*}[t]
  \centering
  \includegraphics[width=.95\textwidth]{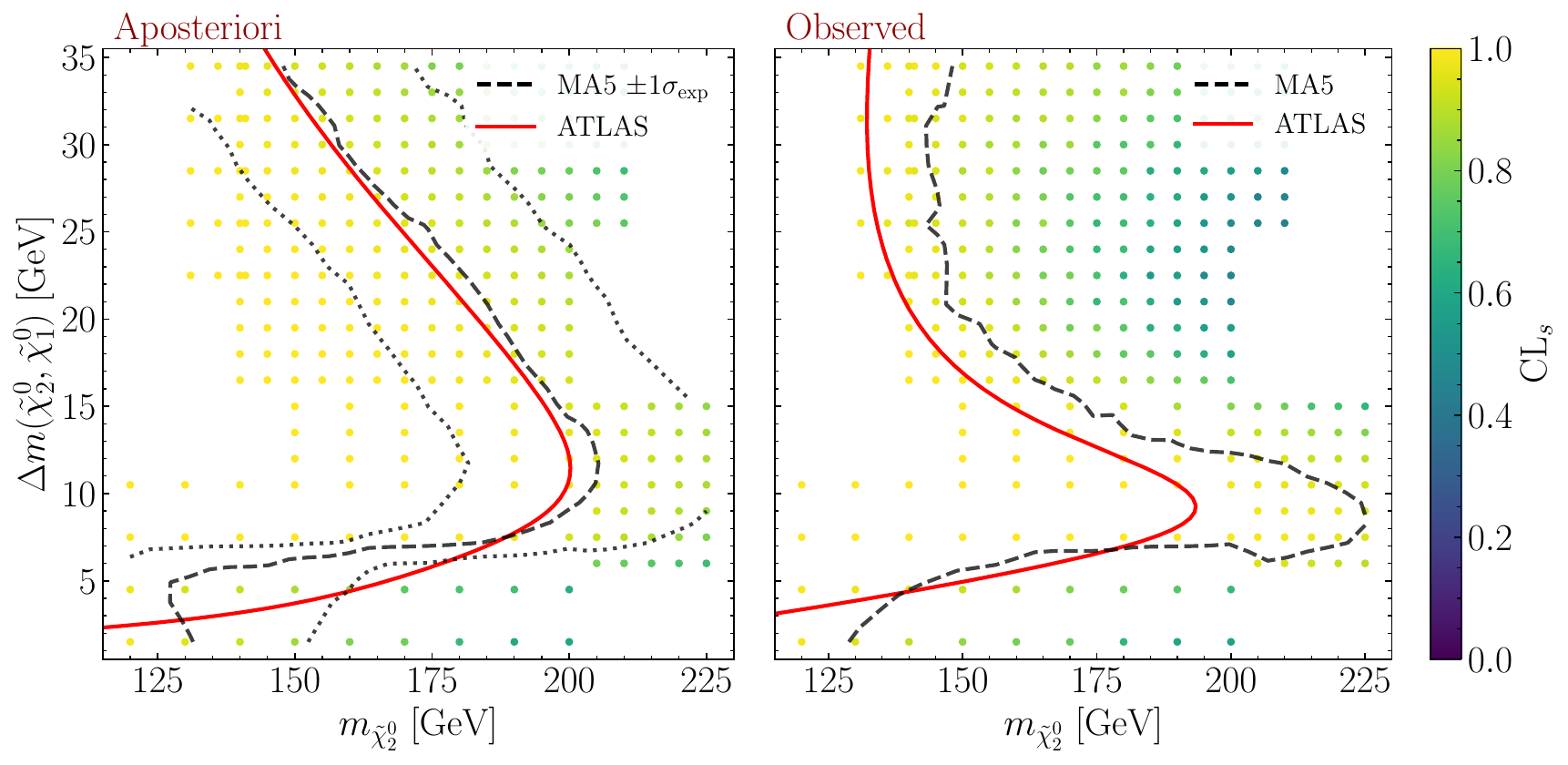}
  \caption{Comparison of \madanalysis\ exclusions with official ATLAS results for the ATLAS-SUSY-2018-16 analysis. We show expected exclusion contours assuming that observations align with the Standard Model (left), as well as observed exclusions based on ATLAS data (right). Results are presented in terms of the mass of the second neutralino, $m_{\tilde\chi_2^0}$, and the mass difference between the two lightest neutralinos, $\Delta m(\tilde\chi_2^0, \tilde\chi_1^0)$.\label{fig:atlas_2018_16_validation}}
\end{figure*}

Given the small signal region selection efficiencies, the validation of the analysis requires large-statistics event samples, which are generated as follows. We consider the three production processes
\begin{equation}
  p p \to \tilde\chi_1^0    \tilde\chi_2^0, \quad
  p p \to \tilde\chi_1^\pm \tilde\chi_2^0, \quad
  p p \to \tilde\chi_1^\pm \tilde\chi_1^\mp,
\label{eq:process}\end{equation}
and treat each of them separately. For every process, an event sample is generated by combining matrix elements including up to two additional partons, using the MLM scheme~\cite{Mangano:2006rw, Alwall:2008qv} with a matching scale set to a quarter of the average mass of the produced electroweakinos. Hard-scattering events are generated by convoluting leading-order (LO) matrix elements obtained with \lstinline{MadGraph5_aMC@NLO}~\cite{Alwall:2014hca} (\lstinline{MG5aMC}) using its built-in MSSM UFO implementation~\cite{Degrande:2011ua, Darme:2023jdn} documented in~\cite{Duhr:2011se}, and the LO set of NNPDF4.0 parton distribution functions~\cite{NNPDF:2021njg, Buckley:2014ana}. Chargino decays are handled through \lstinline{Pythia 8}~\cite{Bierlich:2022pfr} under a flat phase-space assumption, this approximation being justified in the compressed scenarios considered here as the chargino decay products typically enter the analysis primarily through observables which are insensitive to detailed decay angular correlations. In contrast, we employ the decay chain functionality in \lstinline{MG5aMC} to simulate neutralino decays. This ensures that spin correlations are preserved, resulting in a correct di-lepton invariant mass distribution. Moreover, this approach increases the SR selection efficiencies by mitigating the loss of events due to the relatively low (approximately 10\% for the considered scenarios) leptonic branching ratio of the second neutralino. Finally, \lstinline{Pythia 8} is also used for parton showering and hadronisation. After MLM matching, the final event sample relevant for each tested benchmark scenario contains an average of 2.5 million events per process.

We rescale the generated event samples so that the corresponding production cross sections match next-to-leading order predictions, supplemented by next-to-leading logarithmic threshold corrections in the strong coupling (NLO+NLL). We compute these cross sections using \lstinline{Resummino}~\cite{Fuks:2013vua, Fiaschi:2023tkq}, separately for each of the three processes in eq.~\eqref{eq:process}. Finally, exclusion limits are derived using \spey and its interface with \lstinline{pyhf}, and we include a 5\% statistical uncertainty on the signal to remain consistent with the ATLAS limit-setting procedure. 

The validation of our recast implementation is performed for a selected set of benchmark scenarios featuring light higgsinos, the relevant electroweakino states being thus defined as
\begin{equation}
 \tilde \chi_{1,2}^0 \sim -\frac {1}{\sqrt{2}} \left(\mp \tilde H^0_u + \tilde H^0_d \right)\,,\qquad 
 \tilde \chi_{1}^\pm \sim  \tilde H_{u,d}^\pm\,,
\end{equation}
where $\tilde H_{u,d}^{0,\pm}$ denote the neutral and charged higgsino gauge eigenstates associated with the up-type and down-type Higgs doublets. We emphasise that the choice of a real neutralino mixing matrix leads to a negative mass eigenvalue for $m_{\tilde{\chi}^0_2}$, which reflects only a phase convention for the Majorana fields. This does not affect physical observables and is correctly handled by \lstinline{MadGraph5_aMC@NLO} and \lstinline{Pythia 8}. We compare predictions obtained with \madanalysis to those obtained with \lstinline{HackAnalysis} at the level of both signal region yields and efficiencies for definite benchmark scenarios, as well as at the level of the final exclusion contours in the electroweakino mass plane. The \lstinline{HackAnalysis} implementation is thus used here purely as an independent validation reference, given that it has been validated in~\cite{Agin:2024yfs}. Additionally, we compute exclusion contours and compare them with the official ATLAS limits for scenarios in which the masses of the two neutralinos are free parameters, while the mass of the chargino is fixed at the midpoint between them. An excellent agreement is observed in all cases. 

To illustrate the validation procedure, figure~\ref{fig:atlas_2018_16_validation} presents a comparison of the expected and observed exclusion limits for the considered simplified higgsino model. The results are projected onto a two-dimensional plane defined by the mass of the heavier neutralino $m_{\tilde\chi_2^0}$ and its mass difference with the lightest neutralino, 
\be \label{sec:deltam}
  \Delta m(\tilde\chi_2^0, \tilde\chi_1^0) = m_{\tilde\chi_2^0} - m_{\tilde\chi_1^0}\,.
\ee
The left panel of the figure shows the expected exclusion contours, computed under the assumption that observations match SM predictions (\lstinline{spey.ExpectationType.aposteriori} in the \lstinline{spey} terminology). The \madanalysis\ exclusion is depicted as a dashed black line, along with its $1\sigma$ variation, while the official ATLAS exclusion is shown as a solid red line. We obtain an amazing agreement between the ATLAS results and the \madanalysis predictions. The right panel presents the corresponding observed exclusion contours, derived using ATLAS data when initialising the statistical model (\lstinline{spey.ExpectationType.observed} in the \lstinline{spey} terminology). The agreement remains reasonably good, particularly considering the uncertainties inherent to the \madanalysis recast. Note that we do not include control regions in the recast, nor do we include a signal uncertainty, which was found to improve agreement around $\Delta m \sim 10$ GeV \cite{Agin:2024yfs}. 

Given this level of agreement, we consider our recast implementation validated. The corresponding code, which leverages the SFS framework~\cite{Araz:2020lnp} for detector simulation, is publicly available in the \madanalysis\ dataverse at the webpage \href{https://doi.org/10.14428/DVN/1CPXHX}{https://doi.org/10.14428/DVN/1CPXHX}~\cite{1CPXHX_2025}. It is fully integrated within the PAD and is automatically downloaded during installation (achieved by executing the command \lstinline{install PADForSFS} in the \madanalysis\ interpreter).

\subsection{The ATLAS-SUSY-2019-09 search with multiple leptons}\label{sec:atlas_susy_2019_09}
In the ATLAS-SUSY-2019-09 analysis~\cite{ATLAS:2021moa}, the ATLAS Collaboration conducted a search for electroweakino pair production in final states with three leptons and missing transverse energy, using 139~fb$^{-1}$ of proton-proton collision data at $\sqrt{s} = 13$~TeV. The analysis focuses primarily on chargino-neutralino pair production ($pp\to\tilde{\chi}_1^\pm\tilde{\chi}_2^0$), targets scenarios where the produced electroweakinos decay via intermediate electroweak vector and scalar bosons, and includes distinct sets of signal regions optimised for different mass spectra. Three sets of signal regions are hence defined. The `on-shell $WZ$’ regions are dedicated to events where the $\tilde{\chi}_1^\pm$ and $\tilde{\chi}_2^0$ states decay via on-shell $W$ and $Z$ bosons, thus selecting events featuring a di-lepton system with an invariant mass around $m_Z$ and large $\met$. The `off-shell $WZ$’ regions are tailored for moderate and small electroweakino mass splitting. The $W$ and $Z$ bosons emerging from the electroweakino decays are thus virtual and off-shell, which yields softer leptons and a lower amount of $\met$. Finally, the `$Wh$' regions select events in which the decay $\tilde{\chi}_2^0 \to h\tilde{\chi}_1^0$ occurs, exploiting the presence of $b$-jets from Higgs decays in addition to leptons and missing transverse energy.

As this work focuses on excesses relevant to compressed electroweakino mass spectra, we only consider the implementation of the off-shell $WZ$ signal regions where excesses have actually been observed. The preselection begins by requiring exactly three leptons from which we can form at least one OSSF pair and which can be relatively soft. Sensitivity to both soft and harder leptons is ensured by employing a combination of triggers which rely on the presence in the events of either missing transverse energy or one, two, or three leptons with varied $p_T$ thresholds. Additional kinematic and object-based requirements are applied next. The minimum $m_{\ell\ell}^\mathrm{min}$ and maximum $m_{\ell\ell}^\mathrm{max}$ invariant mass of all possible $n_\mathrm{OSSF}$ OSSF lepton pairs that can be formed must fall within predefined values, which effectively implements a veto on the production of a $Z$ boson or hadronic resonances. Further selection criteria include isolation requirements on the transverse separation of all lepton pairs $\Delta R_{3\ell}$, stricter constraints on the leptons forming OSSF pairs (and their minimum separation $\Delta R_{\mathrm{OSSF}}$), and the rejection of events containing reconstructed $b$-jets. 

Following this basic preselection, the analysis defines four categories of signal regions based on the properties of the three leptons, the amount of missing transverse energy, and the number of reconstructed hard jets. Two low-$\met$ categories of signal regions, SR${}^{\text{offWZ}}_{\text{low} \slashed{E}_T}$-0j and SR${}^{\text{offWZ}}_{\text{low} \slashed{E}_T}$-nj, impose an upper bound on the amount of transverse missing energy in the events, requiring it to be below 50~GeV and 200~GeV respectively. In the former class of SRs, events with jets having $p_T > 30$~GeV are additionally vetoed, whereas in the latter class of SRs, at least one such jet is required, effectively increasing the recoil of the electroweakino system and thus yielding more missing transverse energy. Conversely, two class of high-$\met$ signal regions are introduced, SR${}^{\text{offWZ}}_{\text{high} \slashed{E}_T}$-0j and SR${}^{\text{offWZ}}_{\text{high} \slashed{E}_T}$-nj, in order to target events with larger transverse missing energy. Here, we impose minimum $\met$ requirements of 50~GeV and 200~GeV, respectively. In addition to these kinematic selections, the preselection applies an object-based $\met$ significance constraint to account for potential mis-measurement of the properties of the visible reconstructed objects, and it enforces basic lepton $p_T$ cuts and constraints on the invariant mass $m_{3\ell}$ of the tri-lepton system if the lepton associated with the $W$-boson decay $\ell_W$ is an electron. A complete list of criteria is provided in the upper panel of table~\ref{tab:201909_selection}.

\begin{table*}
  \centering\renewcommand{\arraystretch}{1.4}\setlength{\tabcolsep}{12pt}
  \begin{tabular}{l | c c c c}
    Variable & SR${}^{\text{offWZ}}_{\text{low} \slashed{E}_T}$-0j & SR${}^{\text{offWZ}}_{\text{low} \slashed{E}_T}$-nj & SR${}^{\text{offWZ}}_{\text{high} \slashed{E}_T}$-0j & SR${}^{\text{offWZ}}_{\text{high} \slashed{E}_T}$-nj \\
    \hline
    Number of leptons & \multicolumn{4}{c}{3 leptons} \\
    $n_\mathrm{OSSF}$ & \multicolumn{4}{c}{$\geq 1$} \\
    $m_{\ell\ell}^\mathrm{max}$ [GeV] & \multicolumn{4}{c}{$<75$} \\
    $m_{\ell\ell}^\mathrm{min}$ [GeV] & \multicolumn{4}{c}{$\in [1,75]$} \\
    Number of $b$-jets & \multicolumn{4}{c}{$0$ $b$-jets} \\
    $\min \Delta R_{3\ell}$ & \multicolumn{4}{c}{$>0.4$} \\
    Resonance veto [GeV] & \multicolumn{3}{c|}{ $m_{\ell\ell}^\mathrm{min} \notin [3,3.2]$; $m_{\ell\ell}^\mathrm{min} \notin [9,12]$} & $-$ \\
    Trigger & \multicolumn{2}{c|}{(multi-)lepton} & \multicolumn{2}{c}{(multi-)lepton\ or $\met$} \\
    Number of jets with $p_T>30~\text{GeV}$ & $= 0$ & $\geq 1$ & $= 0$ & $\geq 1$\\
    $\met$ [GeV] & $<50$  & $<200$ & $>50$ & $>200$ \\
    $\met$ significance & $>1.5$ & $>3.0$ & $>3.0$ & $>3.0$ \\
    Lepton $p_T$'s [GeV] & \multicolumn{3}{c|}{$>10$} & \ $>4.5$ $(3.0)$ \ for $e(\mu)$ \\
    $|m_{3\ell}-m_Z|$ [GeV] & \multicolumn{2}{c|}{$>20$ (if $\ell_W=e$)} & \multicolumn{2}{c}{$-$} \\
    $\min \Delta R_{\mathrm{OSSF}}$ & \multicolumn{2}{c|}{$[0.6,2.4]$ (if $\ell_W=e$)} & \multicolumn{2}{c}{$-$} \\
  \end{tabular}\vspace{.5cm}
  
  \setlength{\tabcolsep}{10pt}\renewcommand{\arraystretch}{1.5}

  \begin{tabular}{l | c c c c c c c c} 
    \multirow{2}{*}{Variable} & \multicolumn{8}{c}{Selection requirements on the regions SR${}^{\text{offWZ}}_{\text{low} \slashed{E}_T}$} \\
    & {b} & {c} & {d} & {e} & {f1} & {f2} & {g1} & {g2} \\
   $m_{\ell\ell}^\mathrm{min}$ [GeV] & [12, 15] & [15, 20] & [20, 30] & [30, 40] & [40, 60] & [40, 60]  & [60, 75] & [60, 75] \\ \hline
   $m_{\ell\ell}^\mathrm{max}$ [GeV] & $<60$ & $<60$ & $<60$ & $<60$ & $-$ & $-$ & $-$ & $-$ \\
   $m_T^{\text{mllmin}}$ [GeV] & $<50$ & $<50$ & $<50$ & $<60$ & $<60$ & $>90$ & $<60$ & $>90$ \\
   $m_{T2}^{100}$ [GeV] & $<115$ & $<120$ & $<130$  & $-$ & $-$ & $-$ & $-$ & $-$\\
   $\Delta R_{\text{OSSF}}$ & $<1.6$ & $<1.6$ & $<1.6$ & $-$ & $-$ & $-$ & $-$ & $-$\\
   Lepton $p_T$'s [GeV] & $>10$ & $>10$ & $>10$ & $>10$ & $>15$ & $>15$ & $>15$ & $>15$\\ \hdashline
   SR${}^{\text{offWZ}}_{\text{low} \slashed{E}_T}$-0j:\ \ $p_T^\text{lep}|/\met$  & $<1.1$& $<1.1$  & $<1.1$ & $<1.3$ & $<1.4$ & $<1.4$ & $<1.4$ & $<1.4$ \\
   SR${}^{\text{offWZ}}_{\text{low} \slashed{E}_T}$-0j:\ \ $m_{3\ell}$ [GeV] & $-$ & $-$ & $-$ & $-$ & $>100$ & $>100$ & $>100$ & $>100$ \\\hdashline
   SR${}^{\text{offWZ}}_{\text{low} \slashed{E}_T}$-nj : \ \ $p_T^\text{lep}|/\met$ & $<1.0$& $<1.0$  & $<1.0$ & $<1.0$ & $<1.2$ & $<1.2$ & $<1.2$ & $<1.2$ \\
  \end{tabular}\vspace{.5cm}

  \setlength{\tabcolsep}{8pt}
  
  \begin{tabular}{l | c c c c c c c c c} 
   \multirow{2}{*}{Variable} & \multicolumn{9}{c}{Selection requirements on the regions SR${}^{\text{offWZ}}_{\text{high} \slashed{E}_T}$} \\
    & {a} & {b} & {c} & {d} & {e} & {f1} & {f2} & {g1} & {g2} \\
   $m_{\ell\ell}^\mathrm{min}$ [GeV]  & [1, 12] & [12, 15] & [15, 20] & [20, 30] & [30, 40] & [40, 60] & [40, 60]  & [60, 75] & [60, 75] \\ \hline
    $m_{T2}^{100}$ [GeV]  & $<112$ & $<115$ & $<120$ & $<130$ & $<140$ & $<160$ & $<160$ & $<175$ & $<175$ \\
   SR${}^{\text{offWZ}}_{\text{high} \slashed{E}_T}$-0j: \ $p_T(\ell)$ [GeV]& $\times$ & \multicolumn{7}{c}{$ p_T(\ell_1)>25$, $ p_T(\ell_2)>15$, $ p_T(\ell_3)>10$} & \\
   SR${}^{\text{offWZ}}_{\text{high} \slashed{E}_T}$-0j: \ $m_T^{\text{mllmin}}$ [GeV] & $\times$ & $<50$ & $<50$ & $<60$ & $<60$ & $<70$ & $>90$ & $<70$ & $>90$ \\ \hdashline
  SR${}^{\text{offWZ}}_{\text{high} \slashed{E}_T}$-nj: \ $p_T^\text{lep}|/\met$ & $<0.2$ & $<0.2$ & $<0.3$ & $<0.3$ & $<0.3$ & $<1.0$ & $<1.0$ & $<1.0$ &$<1.0$\\[.3cm]
  \end{tabular}\vspace{.2cm}
  \caption{Preselection (upper table) and selection (two lower tables) requirements defining the various off-shell $WZ$ signal regions of the ATLAS-SUSY-2019-09 analysis. The suffixes {a}, {b}, ..., {g1} and {g2} refer to the different exclusive SR constructed in the four main classes of SRs (like for example SR${}^{\text{offWZ}}_{\text{low} \slashed{E}_T}$-0jb, SR${}^{\text{offWZ}}_{\text{low} \slashed{E}_T}$-0jc, \etc).} \label{tab:201909_selection}
\end{table*}

A relatively large number of exclusive SRs are next defined by primarily enforcing non-overlapping requirements on the minimum invariant mass of OSSF lepton pairs, $m_{\ell\ell}^{\text{min}}$. In addition, further selection criteria on the lepton transverse momenta and angular separation are implemented to ensure good reconstruction, as well as constraints on the transverse mass $m_T^{\text{mllmin}}$ of the lepton $\ell_W$ assigned to the $W$ boson and the stransverse mass $m_{T2}^{100}$ of the tri-lepton system computed with a test mass of $100$~GeV~\cite{Lester:1999tx, Cheng:2008hk}. Further refinements are made by imposing additional constraints on the tri-lepton invariant mass $m_{3\ell}$ and on the ratio $p_T^\text{lep}/\met$, where the numerator represents the transverse momentum of the tri-lepton system. This non-standard last requirement is particularly useful to reduce signal contamination from mis-measurement effects. All these selections are listed in the four lower panels of table~\ref{tab:201909_selection}.

\begin{figure*}[t]
  \centering
  \includegraphics[width=.95\textwidth]{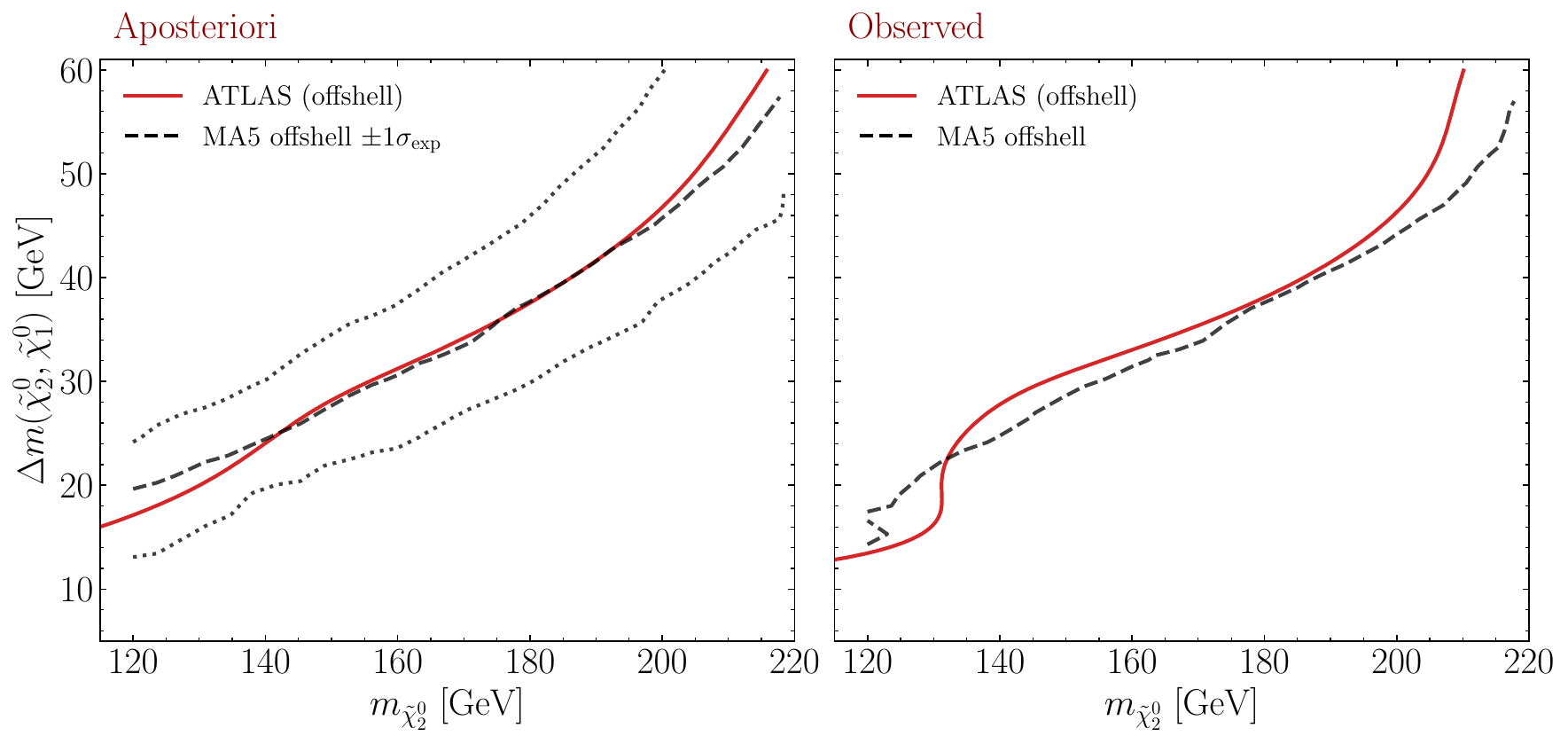}
  \caption{Comparison of \madanalysis\ exclusions (black) with official ATLAS results (red) for the ATLAS-SUSY-2019-09 analysis. We show expected exclusion contours assuming that observations align with the Standard Model (left), as well as observed exclusions based on ATLAS data (right). Results are presented in terms of the mass of the second neutralino $m_{\tilde\chi_2^0}$ and the mass difference between the two lightest neutralinos $\Delta m(\tilde\chi_2^0, \tilde\chi_1^0)$.\label{fig:atlas_2019_09_validation}}
\end{figure*}

The implementation and validation of the ATLAS-SUSY-2019-09 analysis in \madanalysis pose similar challenges to those encountered for the ATLAS-SUSY-2018-16 analysis (see section~\ref{sec:atlas_susy_2018_16}). The recasting of this search is particularly non-trivial due to the large number of observables and triggers involved. To address these complexities, we follow the approach introduced in~\cite{Agin:2024yfs}, which incorporates ATLAS measurements~\cite{ATLAS:2019dpa, ATLAS:2020gty} to model lepton and missing transverse energy triggers, ensuring a faithful reproduction of the original selection efficiencies. A crucial aspect of the analysis is the computation of the $\met$ significance, which is used to suppress backgrounds arising from mis-measured visible objects. This quantity is defined as~\cite{ATLAS:2018uid}
\begin{equation}
   \Big[\mathcal{S}(\met)\Big]^2 = \ptmiss \cdot (V_{xy}^{\rm total} )^{-1} \cdot \ptmiss\,,
\end{equation}
where $V_{xy}^{\rm total}$ represents the (total) covariance matrix of uncertainties in the transverse plane. This matrix is constructed as
\begin{equation}
  V_{xy} = \sum_{\text{objects}} R^{-1} (\varphi) 
  \begin{pmatrix} 
    \sigma_\perp^2 & 0 \\ 
    0 & \sigma_\parallel^2 
  \end{pmatrix} 
  R(\varphi)\,,
\end{equation}
where $R(\varphi)$ is a rotation matrix corresponding to the azimuthal direction $\varphi$ of each reconstructed object in the event. In this formulation, the resolution uncertainties on the four-momentum of each object are decomposed into two components, $\sigma_\parallel$, which accounts for the uncertainty along the direction of motion of the object, and $\sigma_\perp$, which corresponds to the uncertainty in the plane perpendicular to the object's motion. Additionally, a universal diagonal soft term is included to capture contributions from diffuse low-energy activity in the detector.

To validate our implementation, we consider the higgsino simplified model introduced in section~\ref{sec:atlas_susy_2018_16}, in which the chargino mass is fixed to the average of the two lightest neutralino masses. We focus on the associated production of the lightest chargino and the second-lightest neutralino,
\begin{equation}
  p p \to \tilde\chi_1^\pm \tilde\chi_2^0\,,
\end{equation}
as this channel dominates the new-physics production of final states with three leptons and missing transverse energy, with the other supersymmetric contributions remaining relatively negligible. Corresponding events are generated following the strategy described in section~\ref{sec:atlas_susy_2018_16}. Here, we simulate a total of 4,000,000 parton-level events with \lstinline{MG5aMC} prior to MLM merging, utilising the decay chain functionality of the code to handle neutralino decays and thereby ensuring a faithful modelling of spin correlations and lepton kinematics. The merged dataset then comprises approximately 2.5 million events per mass configuration considered in our scan.

Figure~\ref{fig:atlas_2019_09_validation} compares predictions obtained with our recast implementation in \madanalysis, using the SFS framework~\cite{Araz:2020lnp} for detector simulation, with the official ATLAS results. It displays exclusion contours in a two-dimensional plane defined by the mass of the second-lightest neutralino $m_{\tilde \chi^0_2}$ on the $x$-axis and the mass difference $\Delta m(\tilde\chi_2^0, \tilde\chi_1^0)$ defined in eq.~\eqref{sec:deltam} on the $y$-axis. The left panel of the figure considers the expected exclusion under the assumption that observations match SM predictions (\ie\ \lstinline{spey.ExpectationType.aposteriori} in the \lstinline{spey} terminology). It shows the ATLAS exclusion in red and the \madanalysis\ one in black, together with its $1\sigma$ uncertainty band indicated with a dotted line. In addition, the right panel presents the exclusion determined from observed data (\lstinline{spey.ExpectationType.observed}). An excellent agreement is obtained in both cases, which validates our recast implementation. The corresponding analysis code has consequently been released within the \madanalysis dataverse at \href{https://doi.org/10.14428/DVN/LRKDTQ}{https://doi.org/10.14428/DVN/LRKDTQ}~\cite{LRKDTQ_2025}, and is now fully integrated within the PAD.


\section{Singlino-like dark matter in the NMSSM}\label{sec:application}

The ATLAS analyses ATLAS-SUSY-2018-16 and ATLAS-SUSY-2019-09 are of particular interest as they exhibit mild excesses in specific bins of the dilepton invariant mass distribution. In particular, these excesses admit an intriguing interpretation within the MSSM with a spectrum involving light compressed higgsinos. Such an interpretation has addiitonally been shown to be compatible with similar excesses observed in monojet and multijet + $\met$ searches by ATLAS and CMS~\cite{Agin:2023yoq}, and has been further explored in~\cite{Baer:2023olq, Carpenter:2023agq, Altakach:2023tsd, Baer:2023ech, Baer:2024kms, Chakraborti:2024pdn, Martin:2024pxx, Agin:2024yfs, Martin:2024ytt, Agin:2025vgn}. Nevertheless, there is ample motivation to explore alternative scenarios. 

In \cite{Agin:2024yfs}, some of the present authors examined not only the impact of the excesses on MSSM scenarios with a lightest supersymmetric particle (LSP) made of a bino-wino admixture, but also NMSSM scenarios with a singlino-higgsino LSP alongside several non-supersymmetric models. The NMSSM scenario was partially motivated by the desire to simultaneously explain the ATLAS soft-lepton excesses and accommodate a viable dark matter candidate with the correct relic abundance via thermal freeze out. This goal cannot be achieved within the MSSM if the LSP is a light higgsino, as such candidates tend to be underabundant and are strongly constrained by direct-detection experiments.

Concurrently, another group presented an analysis~\cite{Ellwanger:2024vvs} focusing on a slightly different NMSSM scenario in which the mass splitting between the singlino-like LSP $\tilde{S}$ and the higgsino-like next-to-lightest superpartner (NLSP) $\tilde{h}_1$ was restricted to be below 5~GeV. This constraint, absent in the NMSSM study of~\cite{Agin:2024yfs}, was imposed to ensure that the decay $\tilde{h}_1^0 \to \tilde{S} + f\bar{f}$ remains invisibly soft at the LHC. As a result, the electroweakino decay chain effectively terminates at the NLSP, yielding minimal additional missing transverse energy and thereby mimicking an MSSM scenario with a higgsino LSP from the point of view of soft-lepton searches, while retaining a viable dark matter candidate. Notably, the authors of~\cite{Ellwanger:2024vvs} directly compared their results to those of \cite{Agin:2024yfs}, arguing in particular that the compressed spectrum with $\Delta m(\tilde{h}_1^0, \tilde{S}) < 5$~GeV prevents their NMSSM scenario from producing a sizeable monojet signal. They also highlighted the possible existence of a singlino-like scalar Higgs boson with a mass near 95.3~GeV which could account for mild excesses observed in new physics searches in the $\gamma\gamma$ channel at the LHC and in the $b\bar{b}$ final state at LEP~\cite{LEPWorkingGroupforHiggsbosonsearches:2003ing}, though not the LHC excess observed in new resonance searches in the $\tau^+\tau^-$ channel~\cite{ATLAS:2020zms}. Now equipped with validated implementations of the relevant soft-lepton analyses in \madanalysis, we revisit and scrutinise these interesting and well-motivated claims through our own statistical analysis based on full event generation.

\begin{table}
\centering\renewcommand{\arraystretch}{1.4}\setlength{\tabcolsep}{12pt}
\begin{tabular}{l | l }
  Parameter & Input Range\\
\hline
$\lambda$  & $[0.013, 0.019]$\\
$\kappa$  & $[0.0058, 0.0086]$ \\
$A_{\lambda}$ [GeV] & $[-5000, -1820]$\\
$A_{\kappa}$ [GeV] & $[93, 362]$ \\
$\tan\beta$ & $[3.61, 10.9]$\\
$\mu_{\rm eff}$ [GeV] & $[-244, -148]$\\
$M_1$ [GeV] & $[178, 290]$\\
$M_2$ [GeV] & $[304, 5000]$\\
$M_3$ [GeV] & $[739, 5000]$\\
$M_{Q_3}$ [GeV]  & $[272, 5000]$\\
$M_{U_3}$ [GeV] & $[570, 5000]$\\
$A_t$ [GeV] & $[-5000, 142]$
\end{tabular}
\caption{\label{tab:scanInputs}Parameters allowed to vary in our scan over the NMSSM parameter space, following the analysis in \cite{Ellwanger:2024vvs} targeting electroweakino excesses.}
\end{table}

The analysis of~\cite{Ellwanger:2024vvs} presents a scan over the NMSSM parameter space with input ranges summarised in table~\ref{tab:scanInputs}. The authors found that the dark matter relic abundance as measured by the Planck Collaboration from the cosmic microwave background, $\Omega h^2 \approx 0.12$~\cite{Planck:2018vyg}, can be readily satisfied in scenarios with compressed singlino-higgsino spectra primarily via coannihilations. They then identified many benchmark points featuring a singlino-like LSP separated by no more than 5~GeV from the lightest higgsino, along with a singlet-like scalar Higgs boson with a mass near 95.3~GeV. To validate these findings and illustrate our implementation of the ATLAS soft-lepton analyses in \madanalysis, we perform a dedicated scan that closely follows the setup of~\cite{Ellwanger:2024vvs}. Naturally, unless the results are filtered to retain only compressed electroweakino spectra, the scan yields a broader set of benchmark points with singlino-higgsino mass splittings exceeding 5~GeV, \ie\ a set of points distributed across the $(m_{\tilde{h}_1^0}, \Delta m(\tilde{h}_1^0, \tilde{S}))$ plane. Additionally, many points exhibit a significant bino admixture in the NLSP, and some points even feature a bino-like LSP. We find it instructive to first \emph{relax} the 5~GeV splitting condition, allowing us to examine the sensitivity of the ATLAS electroweakino searches to a wider region of the parameter space and to enable a direct comparison with the NMSSM scenario explored in~\cite{Agin:2024yfs}. We nevertheless subsequently return to the tightly constrained scenario of~\cite{Ellwanger:2024vvs} to further investigate its phenomenological implications.

For this purpose, we use the NMSSM implementation provided in version 4.15.2 of \lstinline{SARAH}~\cite{Staub:2008uz, Staub:2013tta, Goodsell:2014bna, Goodsell:2017pdq}, which we employ to generate a Fortran code using routines from the \lstinline{SPheno} library~\cite{Porod:2003um, Porod:2011nf}. Fermion masses are computed including one-loop corrections, while scalar masses include two-loop corrections~\cite{Goodsell:2014pla, Goodsell:2015ira, Braathen:2017izn} and all decay widths are calculated at leading order. In addition, we generate LO model files compatible with \lstinline{CalcHEP}~\cite{Belyaev:2012qa} (and hence with \lstinline{micrOMEGAs}~\cite{Belanger:2010pz, Belanger:2013oya, Alguero:2023zol}), as well as with the \lstinline{MG5aMC} event generator. The tool \lstinline{BSMArt}~\cite{Goodsell:2023iac} is used to orchestrate the scan, calling the above software packages and sampling the NMSSM parameter space using random input values within the ranges indicated in table~\ref{tab:scanInputs}. The final output consists of a set of 235 benchmark points with various electroweakino mixings and mass splittings.

\begin{figure}[t]
  \centering
  \includegraphics[width=.99\columnwidth]{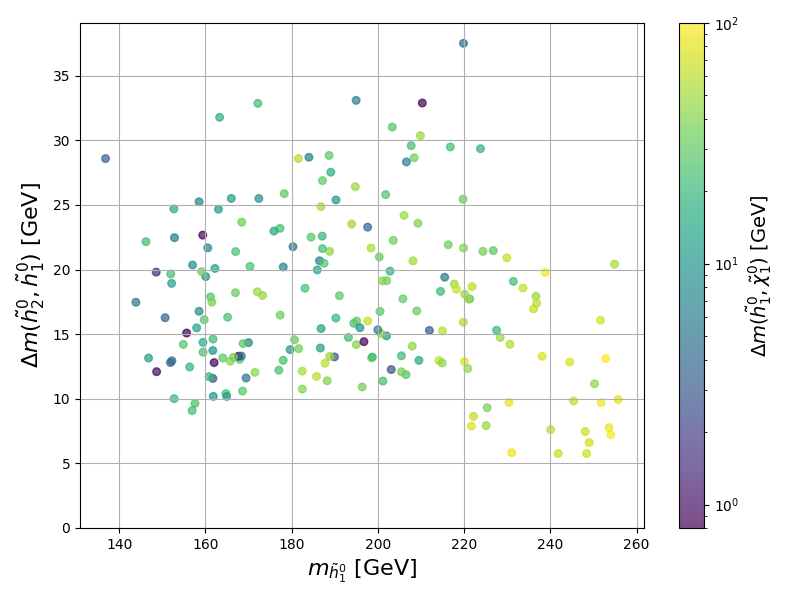}
  \caption{Higgsino spectrum properties as obtained from our scan of the NMSSM parameter space. Results are presented in terms of the mass $m_{\tilde h_1^0}$ of the lightest higgsino-like electroweakino, and the mass difference between the two higgsino-like states $\Delta m(\tilde h_2^0, \tilde h_1^0)$. The colour code indicates the mass difference between the lightest higgsino-like state and the LSP.\label{fig:nmssm_masses}}
\end{figure} 

As the wino-like states are assumed decoupled, we focus on scenarios in which four neutralinos and one chargino are light, with masses typically around a few hundred GeV. In most of these scenarios, the lightest neutralino is singlino-like, followed by higgsino-like and bino-like states. The bino is either the heaviest of the electroweakinos or only slightly lighter than the higgsino. To illustrate the structure of the spectrum in the 235 viable parameter points explored in our scan, we present in figure~\ref{fig:nmssm_masses} the relation between the mass of the lightest higgsino-like neutralino, denoted generically by $\tilde h_1^0$, and the mass splitting $\Delta m(\tilde h_2^0, \tilde h_1^0)$ between the two higgsino-like states. Moreover, the colour code describes the splitting $\Delta m(\tilde{h}_1^0, \tilde{\chi}_1^0)$ between the higgsino-like states and the LSP denoted by $\tilde{\chi}_1^0$. In our convention, we define a \emph{higgsino-like state} as an electroweakino $\tilde\chi_i^0$ with mixing properties such that $|N_{i3}|^2 +|N_{i4}|^2 > |N_{i1}|^2 + |N_{i2}|^2 +|N_{i5}|^2$ (where $N_{ij}$ denotes the $ij$ element of the neutralino mixing matrix). Figure~\ref{fig:nmssm_masses} reveals that both higgsino-like neutralinos are typically light, with masses in the range of 150--250~GeV, and that the higgsino spectrum is compressed, featuring mass splittings of 5--35~GeV. In addition, the charged higgsino is found to lie in the same mass range, completing the nearly degenerate higgsino triplet that turns to lie not too far in mass from the LSP, generally by 1--20~GeV. As discussed in \cite{Agin:2024yfs}, such compressed spectra could lead to a leptons + $\met$ signal emerging from higgsino production and decays that potentially compatible with the observed excesses in the ATLAS-SUSY-2018-16 and ATLAS-SUSY-2019-09 searches.

In agreement with~\cite{Ellwanger:2024vvs}, we find that LSP-NLSP mass splittings in a range of 1--3~GeV correlate well with the correct dark matter relic abundance. Interestingly, we also identify a non-negligible subset of points that are disfavoured by dark matter direct detection (DD) constraints, although only one benchmark point simultaneously satisfies $\Omega h^2 \approx 0.12$ and yields a DD $p$-value below 0.05. Without tailoring the scan to optimise any particular likelihood (though we return to this point below), we find that 18 of the 235 points feature a light scalar state with properties such that the signal strength $\mu_{\gamma\gamma}$ lies within $2\sigma$ of the LHC-preferred value and $\mu_{b\bar{b}}$ lies within $1\sigma$ of the LEP-preferred value. These results thus corroborate the conclusion of~\cite{Ellwanger:2024vvs} that a 95-GeV scalar with suitable diphoton and $b\bar{b}$ couplings emerges naturally in the considered region of the NMSSM parameter space with compressed electroweakinos, and could potentially provide an explanation to excesses appearing in new resonance searches in the $\gamma\gamma$ and $b\bar b$ channels.

To explore the precise scenario proposed in~\cite{Ellwanger:2024vvs}, including the prescribed tiny mass splittings between the singlino and the higgsinos, we supplement the above sample of benchmark scenarios with a focused set of 18 points generated via an efficient two-step workflow in \lstinline{BSMArt}. Specifically, we first perform a fast scan in which only the particle spectrum is computed, fixing $M_{Q_3} = M_{U_3}$ and using the `fitting’ feature of the code to determine the value that reproduces the observed Higgs boson mass. Simultaneously, we employ multiple cores to scan over all other parameters listed in table~\ref{tab:scanInputs}, following a Markov Chain Monte Carlo algorithm guided by a user-defined likelihood function
\begin{align}\mathcal{L} =&\exp\!\left[\frac{(|m_{\tilde{\chi}_{2}^{0}} |\!-\!|m_{\tilde{\chi}_1^0}|\!-\!5)^2}{3}  \!+\! \frac{(|m_{\tilde{\chi}_3^0}|\!-\!|m_{\tilde{\chi}_2^0}|\!-\!20)^2}{10}  \right]. \end{align}
This preferentially selects points consistent with the electroweakino spectrum described in~\cite{Ellwanger:2024vvs}. From the large number of points generated, we select those with sufficiently high likelihood and use their input parameters in a \texttt{read\textunderscore csv} scan in \lstinline{BSMArt}, this time producing the full set of outputs and running the complete toolchain including checks against all relevant constraints. In principle, this procedure can be extended to generate arbitrarily large numbers of viable points in a short time (and could potentially benefit from machine-learning techniques to further automate the workflow). The 18 points produced in this way could thus be seen as representing a slice of the parameter space most compatible with the mass spectrum and assumptions of~\cite{Ellwanger:2024vvs}.

To assess the strength of existing LHC constraints on the benchmark points obtained from our scans, we rely on the simulation toolchain described in section~\ref{sec:atlas_susy_2018_16} to generate NMSSM signal events characterised by missing transverse energy accompanied by jets and/or potentially soft leptons. For each benchmark, we simulate three sets of 2,000,000 events (prior to MLM merging), covering each all production processes involving any pair of the five lighter electroweakinos,
\begin{equation}
  p p \to \tilde\chi_i^0\tilde\chi_j^0\,, \quad
  p p \to \tilde\chi_i^0\tilde\chi_1^\pm\,, \quad
  p p \to \tilde\chi_1^+\tilde\chi_1^-\,,
\end{equation}
with $i,j = 1,\ldots,4$. Event generation is performed using \lstinline{MG5aMC}, combining matrix elements including up to two additional partons, and using the NMSSM UFO model produced with \lstinline{SARAH}. The first two sets of samples focus on a signal featuring missing energy and soft leptons. Because a precise description of the lepton angular distributions is essential, we combine production matrix elements with neutralino decay matrix elements obtained with \lstinline{MG5aMC}, similarly to what was done in~\cite{Agin:2024yfs}. However, due to the richer electroweakino spectrum of the NMSSM, the decay chains are more involved. In the first sample set, we simulate processes in which the heavier neutralinos decay directly into the lightest neutralino and leptons. This sample is dominated by $\tilde{\chi}_1^\pm \tilde{\chi}_2^0$ production with the subsequent decay $\tilde{\chi}_2^0 \to \tilde{\chi}_1^0 \ell \ell$, yielding events possibly featuring additional initial-state-radiation jets. Chargino decays are instead handled by \lstinline{Pythia 8}. The second sample set specifically targets the decay $\tilde{\chi}_3^0 \rightarrow \tilde{\chi}_2^0 \ell \ell$. In our scan, both the $\tilde{\chi}_2^0$ and $\tilde{\chi}_3^0$ states are predominantly higgsino-like, so the dominant production modes involve $\tilde{\chi}_2^0 \tilde{\chi}_3^0$ and $\tilde{\chi}_3^0 \tilde{\chi}_1^\pm$ pairs, again possibly with additional jets from initial-state radiation. Ultimately, samples from these first two sets are combined in proportion to their production cross sections. Finally, the third set of samples is inclusive in the electroweakino decays, which are thus performed entirely in \lstinline{Pythia 8}. As a result, this sample naturally includes final states enriched in jets, since the branching fractions of heavier electroweakinos into $\tilde{\chi}_1^0$ and soft hadronic activity are sizeable. 

We use the leptonic samples to derive constraints from the ATLAS-SUSY-2018-16 and ATLAS-SUSY-2019-09 analyses. As for the validation of the implementation of these two analyses in \madanalysis, the explicit neutralino leptonic decays at the event-generation level ensure a sufficiently large number of signal events (hence good statistics) passing all selection cuts. Meanwhile, we use the jet-enriched sample to test compatibility with LHC searches targeting final states with jets and missing transverse energy that also exhibit mild excesses in data. Specifically, we recast the ATLAS-EXOT-2018-06~\cite{ATLAS:2021kxv}, ATLAS-SUSY-2018-22~\cite{ATLAS:2020syg}, CMS-EXO-20-004~\cite{CMS:2021far} and CMS-SUS-19-006~\cite{CMS:2019zmd} analyses. These four searches collectively probe a variety of jets + $\met$ signatures with different jet multiplicities and hardness. The two CMS analyses and the ATLAS-EXOT-2018-06 search are available publicly via the \madanalysis\ PAD~\cite{REPAMM_2023, IRF7ZL_2021, 4DEJQM_2020, Arina:2025zpi, CMS:2021far, Fuks:2021zbm, Mrowietz:2020ztq}, whereas the implementation and validation of the ATLAS-SUSY-2018-22 search has just been released~\cite{Agin:2025vgn, DVN/F65KMF_2025}.

\subsection{Constraints from soft leptons and missing energy}
\begin{figure*}\centering
  \includegraphics[width=0.45\textwidth]{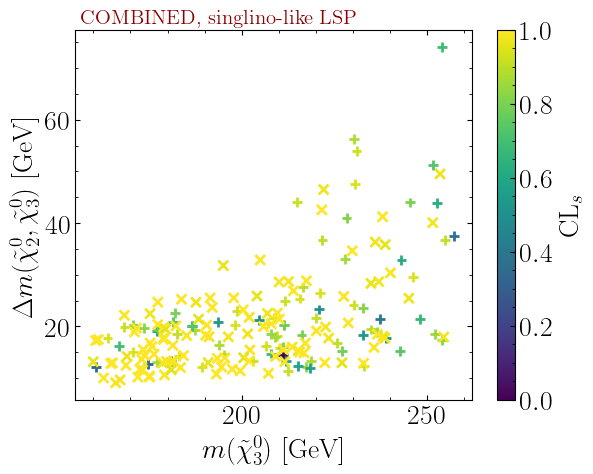}\hfill
  \includegraphics[width=0.45\textwidth]{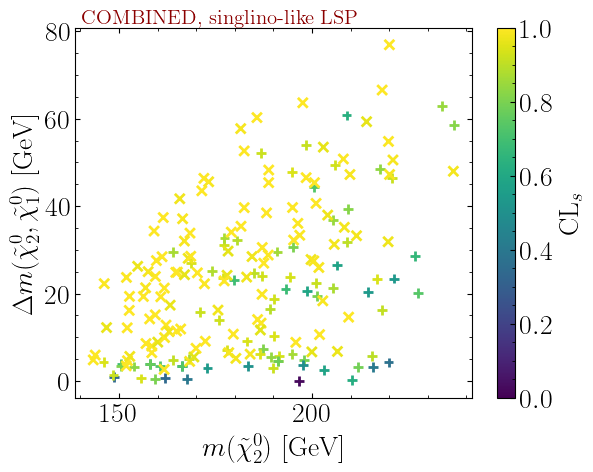}
  \caption{\label{FIG:NMSSM23v3} Combined soft lepton exclusion limits for the considered NMSSM scenario, presented in the $(m(\tilde{\chi}_3^0), \Delta m (\tilde{\chi}_3^0,\tilde{\chi}_2^0))$ [left] and $(m(\tilde{\chi}_2^0), \Delta m (\tilde{\chi}_2^0,\tilde{\chi}_1^0))$ [right] planes.}\vspace{.4cm}
  \includegraphics[width=0.45\textwidth]{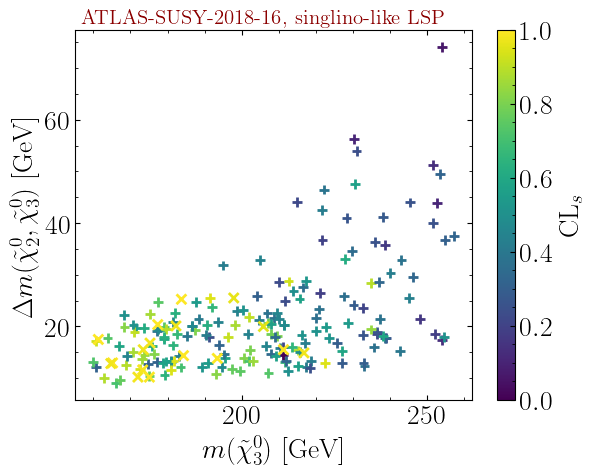}\hfill
  \includegraphics[width=0.45\textwidth]{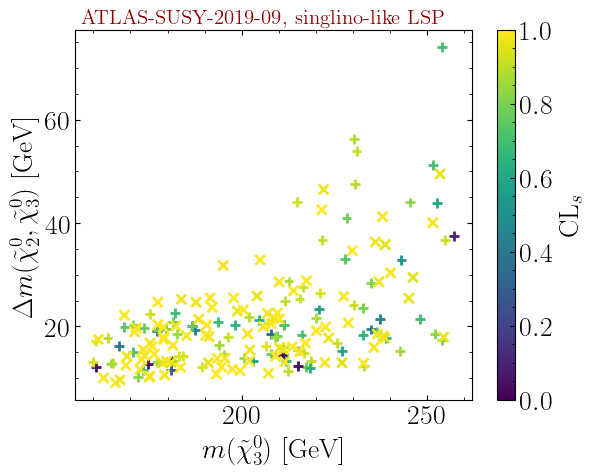}
\caption{\label{FIG:NMSSM2leptons} Exclusion limits in the NMSSM scenario from ATLAS-SUSY-2018-16 (left) and ATLAS-SUSY-2019-09 (right) in the $(m(\tilde{\chi}_3^0), \Delta m (\tilde{\chi}_3^0,\tilde{\chi}_2^0))$ plane.}
\end{figure*}

The hypothesis put forward in \cite{Ellwanger:2024vvs} was that the soft-lepton signal predicted in the considered NMSSM scenario would closely resemble that expected from higgsinos in the MSSM. Here, the $\tilde{\chi}_2^0$, $\tilde{\chi}_3^0$ and $\tilde{\chi}_1^\pm$ states are largely higgsino-like, and the decays of the $\tilde{\chi}_3^0$ and $\tilde{\chi}_1^\pm$ electroweakinos proceed predominantly to $\tilde{\chi}_2^0$ and SM fermions via off-shell $Z$ and $W$ bosons, respectively. It is therefore reasonable to expect that both the exclusion limits and the potential alignment with excesses could be represented in the $(m(\tilde{\chi}_3^0), \Delta m (\tilde{\chi}_3^0,\tilde{\chi}_2^0))$ plane, allowing direct comparison with the ATLAS and CMS results from conventional higgsino searches. In the left panel of figure~\ref{FIG:NMSSM23v3}, we show the outcome of our scan in this plane, combining the limits from the two ATLAS soft-lepton analyses. We present our results as a scatter plot, which clearly reveals the absence of any simple pattern: a substantial number of points are excluded even though they would evade constraints in the corresponding MSSM higgsino scenario, this fact holding also at relatively large neutralino masses and mass splittings.

To better understand this behaviour, we additionally present the results in the $(m(\tilde{\chi}_2^0), \Delta m(\tilde{\chi}_2^0,\tilde{\chi}_1^0))$ plane in the right panel of figure~\ref{FIG:NMSSM23v3}. This view makes it evident that the mass splitting between the higgsino-like states and the singlino-like LSP plays a crucial role: all points with very small splittings remain unexcluded. Further insight comes from examining the exclusion limits from the $2\ell$ and $3\ell$ signal regions separately in the left and right panels of figure~\ref{FIG:NMSSM2leptons}, respectively. The former resembles what one would expect in a higgsino-like scenario, but the latter shows significantly stronger exclusion power, even at small splittings. This increased sensitivity can be traced to the production of $\tilde{\chi}_2^0 \tilde{\chi}_1^\pm$ pairs, whose cross section is large enough to yield an appreciable number of $3\ell$ events thus leading to many excluded benchmark points.

\begin{figure}
\includegraphics[width=0.45\textwidth]{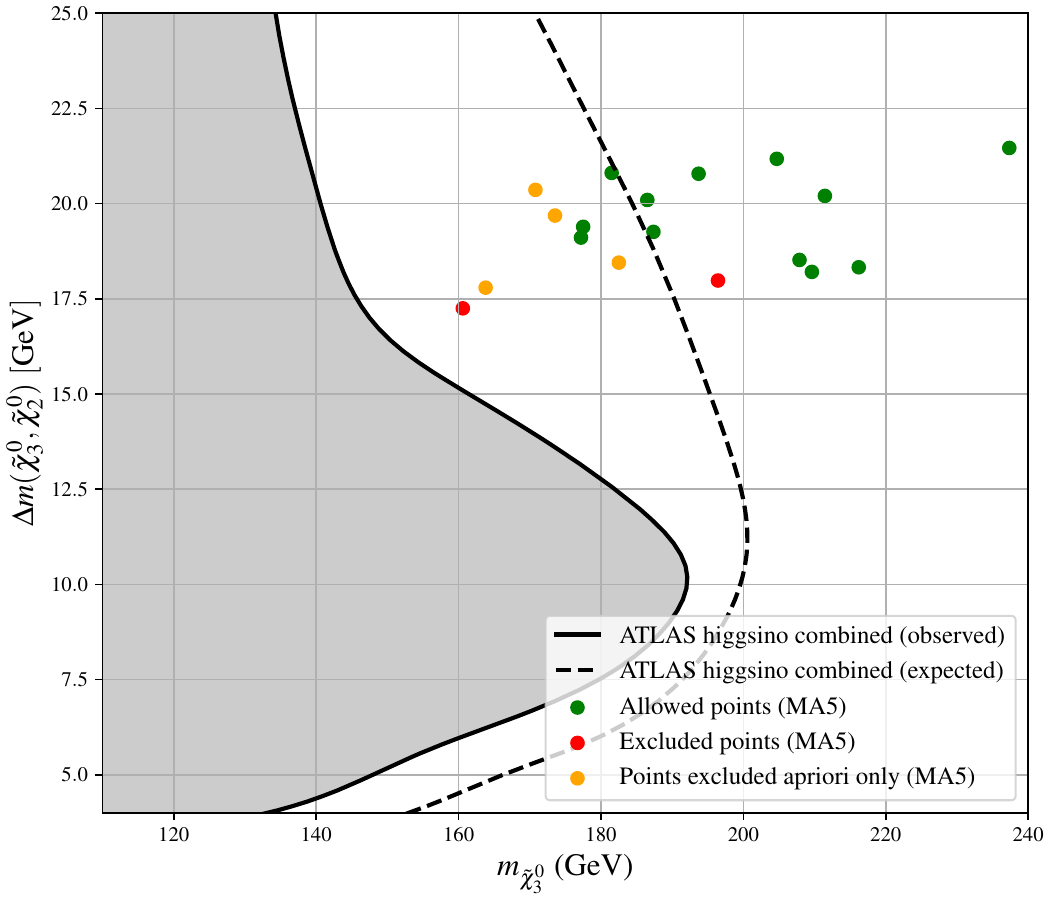}
\caption{\label{FIG:NMSSMrestricted}Exclusion status of selected NMSSM scan points under the combined ATLAS-SUSY-2018-16 and ATLAS-SUSY-2019-09 analyses, shown in the $(m(\tilde{\chi}_3^0), \Delta m (\tilde{\chi}_3^0,\tilde{\chi}_2^0))$ plane. Green points are allowed by the searches; red points are excluded; orange points would be expected to be excluded \emph{a priori} but are in fact allowed by observations. The black curve indicates the ATLAS exclusion contour for the equivalent simplified higgsino scenario.}
\end{figure}

Restricting our attention to the 18 focused scan points, we obtain the scatter plot shown in figure~\ref{FIG:NMSSMrestricted}, which allows for a direct comparison to the ATLAS limits derived under the assumption of a simplified higgsino scenario. These points are particularly relevant because they most closely match the configuration proposed in \cite{Ellwanger:2024vvs}, especially in the region where an excess could be expected. While most of the selected benchmarks behave as anticipated, two of them are unexpectedly excluded. Moreover, it is worth recalling that we use LO cross sections which typically underestimate signal rates compared to NLO+NLL predictions. Our limits should therefore be considered as more conservative. This makes the exclusion of the benchmark scenario at $(m(\tilde{\chi}_3^0), \Delta m(\tilde{\chi}_3^0, \tilde{\chi}_2^0)) = (196.4, 18)$~GeV particularly striking. The electroweakino mass spectrum for this point is given by
\begin{equation}\begin{split}
    &m(\tilde{\chi}_1^0)=172.0~\text{GeV},\quad
    m(\tilde{\chi}_2^0)=178.4~\text{GeV},\\
    &m(\tilde{\chi}_1^\pm)=190.3~\text{GeV}
\end{split}\end{equation}
and the LO cross sections for the relevant production channels are
\begin{align}
\nonumber \sigma(pp \to \tilde{\chi}_3^0 \tilde{\chi}_1^\pm) &=  0.37~\mathrm{pb},\\
\nonumber \sigma(pp \to\tilde{\chi}_2^0 \tilde{\chi}_1^\pm) &= 0.41~\mathrm{pb},\\
\nonumber \sigma(pp \to\tilde{\chi}_1^\pm \tilde{\chi}_1^\mp) &= 0.23~\mathrm{pb},\\
\sigma(pp \to \tilde{\chi}_2^0 \tilde{\chi}_3^0) &=  0.20~\mathrm{pb}.
\end{align}
The leptonic branching ratios for $\tilde{\chi}_3^0 \to \tilde{\chi}_2^0 \ell\ell$ and $\tilde{\chi}_2^0 \to \tilde{\chi}_1^0 \ell\ell$ are each approximately $8.6\%$. What drives the exclusion is then the unexpectedly high event rate in `bin~a' of the SR${}^{\text{offWZ}}_{\text{high}\slashed{E}_T}$ signal region in the ATLAS-SUSY-2019-09 analysis, corresponding to $m_{\ell\ell} \in [1,12]$~GeV. This bin receives significant contributions from the two soft-lepton samples that we generated, \ie\ the one with forced $\tilde{\chi}_3^0$ decays into the second neutralino $\tilde{\chi}_2^0$ and the one in which it decays directly to the LSP. Despite the tight compression between the two higgsino-like states $\tilde{\chi}_2^0$ and $\tilde{\chi}_3^0$ and the singlino-like LSP $\tilde{\chi}_1^0$, both contributions yield a comparable and non-negligible number of signal events in bin~a. As a result, the total yield in this bin is at least twice as large as that expected in the simplified higgsino model, leading not only to a poorer fit to the excess but ultimately to the exclusion of this benchmark scenario.

We can also quantify the degradation of the fit by considering the quantity
\begin{align}
\Delta\chi^2 \equiv  - 2 \log \frac{\mathcal{L}(1, \hat{\hat{\theta}}(1))}{\mathcal{L}(0,\hat{\hat{\theta}}(0))},
\end{align}
where $\mathcal{L}(\mu, \theta)$ is the likelihood of the data given a signal strength $\mu$ and a set of nuisance parameters $\theta$, and where $\hat{\hat{\theta}}(\mu)$ denotes the values of the nuisance parameters that maximise the likelihood for the fixed signal strength $\mu$. A negative value $\Delta\chi^2<0$ thus corresponds to the signal-plus-background hypothesis being favoured over the background-only hypothesis by the data. Some of the authors have used an approximate Bayes factor $B_{10} = \exp(-\Delta\chi^2/2)$ elsewhere~\cite{Goodsell:2024aig,Fuks:2024qdt}, following~\cite{Fowlie:2024dgj}, to measure the strength of this preference. In our large sample of 235 points, we find modest negative values which indicate a mild preference rather than decisive evidence for the signal, owing to significant positive contributions to $\Delta\chi^2$ from the ATLAS-SUSY-2019-09 search. Even within our focused sample of 18 points, only a handful satisfy $\Delta\chi^2 < 0$. Notably, however, this smaller set includes a point with $\Delta\chi^2 = -1.16$ from the combined soft-lepton searches, or equivalently $B_{10} = 1.8$. The associated spectrum is
\begin{align}
    \nonumber m_{\tilde{\chi}_1^0} &= 170.0~\text{GeV},\ m_{\tilde{\chi}_2^0} = \nonumber 172.9~\text{GeV},\\ m_{\tilde{\chi}_3^0} &= 193.7~\text{GeV},\ m_{\tilde{\chi}_1^\pm}= 188.9~\text{GeV}.
\end{align}
The obtained value of $B_{10}$ is smaller than but comparable to similar values found in an MSSM wino-bino (decoupled-higgsino) scenario~\cite{Agin:2025vgn}. We conclude that while a more exhaustive scan of the parameter space focusing on \emph{even smaller} mass splittings between the first two neutralinos may yield better fits to the data, this likely comes at the cost of increased mixing between the singlino and the higgsino, thereby making it harder to satisfy dark matter constraints. The explanation of the soft-lepton excesses in this model appears thus less successful than in simpler MSSM scenarios due to the larger number of expected three-lepton events.

\subsection{Constraints from jets and missing energy}

\begin{figure}
\includegraphics[width=0.45\textwidth]{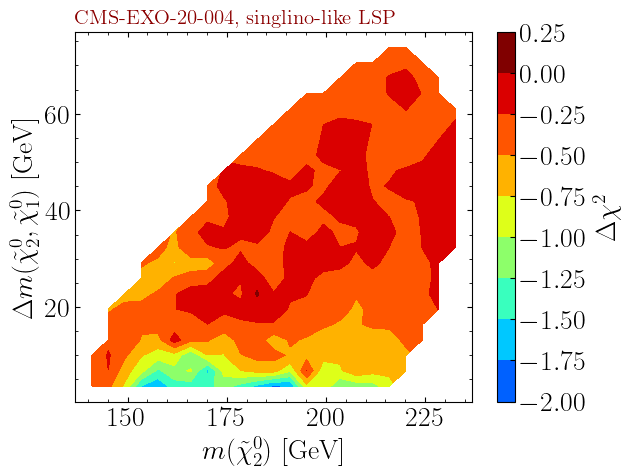}
\caption{\label{FIG:NMSSMmonojet}$\Delta\chi^2$ values from the CMS-EXO-20-004 analysis for the considered NMSSM scenarios, shown in the $(m(\tilde{\chi}_2^0), \Delta m(\tilde{\chi}_2^0, \tilde{\chi}_1^0))$ plane.}
\end{figure}

We then explore the impact of the ATLAS-EXOT-2018-06, CMS-EXO-20-004, ATLAS-SUSY-2018-22 and CMS-SUS-19-006 searches involving jets and missing energy. The first two are typically classified as `monojet' searches, involving low jet multiplicities, and both exhibit substantial excesses (see \cite{Agin:2023yoq} for a discussion). The latter two are `multijet' + $\met$ searches, requiring larger jet multiplicities and targeting supersymmetric signals from coloured particle production with two-step decay chains. We confirm that none of these searches excludes any points in the scanned region of the NMSSM parameter space. However, by computing the associated $\Delta\chi^2$ values (as in the previous section), we can identify the regions of the parameter space that provide a better fit to the data compared to the SM. These results are shown in figure~\ref{FIG:NMSSMmonojet}. 

While the overall fit is poorer than that obtained in simpler models such as the wino-bino scenario considered in \cite{Agin:2025vgn}, small mass splittings still lead to substantial improvements over the SM. For instance, our best-fit point from the previous section yields $\Delta\chi^2 = -1.24$ for CMS-EXO-20-004, corresponding to a combined (soft-lepton + monojet) $\Delta\chi^2 = -2.4$ or $B_{10} = 3.3$. This point therefore represents a rare case where both monojet and soft-lepton excesses can be simultaneously explained within the same model. At present, however, it is not possible to compute the equivalent likelihood information for ATLAS-EXOT-2018-06 due to the absence of a validated \lstinline{pyhf} model; the ATLAS convenors are currently in the process of providing one.

\section{Summary and outlook}\label{sec:conclusion}
We have presented version 1.11 of the \madanalysis software package, which features important new capabilities particularly suited to the reinterpretation of electroweakino searches at the LHC. We described the implementation and validation of two such searches, and applied these, alongside recasts of several monojet and multijet + $\met$ analyses, to a motivated NMSSM scenario~\cite{Ellwanger:2024vvs} which was proposed to simultaneously explain excesses observed in electroweakino and monojet searches, as well as the hint of a Higgs-like boson with mass near 95~GeV. Our analysis reveals that while certain regions of the NMSSM parameter space can accommodate these excesses, especially for small mass splittings in the electroweakino sector, the overall fit is poorer due to the large number of three-lepton events predicted in such scenarios. As a result, a significant portion of the parameter space originally considered viable turns out to be excluded. These findings highlight the importance of full and statistically robust recast analyses when testing complex new physics scenarios.

Looking ahead, we plan to incorporate the CMS soft-lepton search CMS-SUS-18-004~\cite{CMS:2021edw, Agin:2025vgn} into \madanalysis. Once a validated \lstinline{pyhf} statistical model becomes available for the ATLAS jets + missing energy search ATLAS-EXOT-2018-06, we will be able to combine all relevant analyses to conduct a comprehensive study of the compressed electroweak partner sector across different models. Additionally, we aim to extend the statistical capabilities of \madanalysis to include full statistical models from CMS via the \lstinline{Higgs-Combine} framework~\cite{CMS:2024onh}, enabling the use of full ATLAS and CMS likelihoods to better assess the reach of individual analyses.

The mild but persistent excesses in several Run 2 searches underscore the potential importance of these developments. Should they be confirmed with Run 3 data, the ability to rigorously test candidate models will be essential. We will continue to advance our analysis framework to support this effort while we strive to identify new-physics scenarios that offer improved fits over the Standard Model.

\section*{Acknowledgements}

B.~F. and M.~D.~G. are supported in part by Grant ANR-21-CE31-0013, Project DMwithLLPatLHC, from the \emph{Agence Nationale de la Recherche} (ANR), France.


\bibliographystyle{JHEP}
\bibliography{ewkino_recast}

\end{document}